\documentclass[useAMS,usenatbib]{mnras}

\usepackage     [utf8]                  {inputenc}
\usepackage     [T1]                    {fontenc}
\usepackage                             {color}
\usepackage                             {amsmath}
\usepackage     [english]               {babel}
\usepackage                             {natbib}
\usepackage                             {hyperref}

\definecolor{darkblue}{rgb}{0.0,0.0,0.4}
\definecolor{darkgreen}{rgb}{0.0,0.4,0.0}
\hypersetup{
    colorlinks,
    linkcolor=black,
    citecolor=darkgreen,
    urlcolor=darkblue
}
\usepackage{subcaption}
\usepackage{pslatex}
\usepackage{physics}
\usepackage{amsfonts}
\usepackage{url}
\usepackage{graphicx}
\usepackage{amssymb}
\usepackage{textcomp}
\usepackage{soul}
\usepackage[table,x11names]{xcolor}
\usepackage{array}
\usepackage{enumerate}
\usepackage{enumitem}
\usepackage{float}
\usepackage{bm}
\usepackage{empheq}
 \graphicspath{{images/}}
\definecolor{lightgray}{gray}{0.9}
\definecolor{llg}{gray}{0.4}

\newcommand{\bfom}{\bm{\Omega}}
\newcommand{\di}{\mathrm{d}}

\newcommand{\degree}{\ensuremath{^\circ}}

\newcommand{\bfR}{\mathbf{R}}
\newcommand{\bfv}{\mathbf{v}}
\newcommand{\bfG}{\mathbf{G}}

\newcommand{\pc}{\,\mbox{pc}}
\newcommand{\kpc}{\,{ \mbox{kpc}}}
\newcommand{\Myr}{\,{\rm Myr}}
\newcommand{\Gyr}{\,{\rm Gyr}}
\newcommand{\kms}{\,{\mbox{km}\, \mbox{s}^{-1}}}
\newcommand{\de}[2]{\frac{\partial #1}{\partial {#2}}}
\newcommand{\cs}{c_{\rm s}}

\newcommand{\Omegap}{\Omega_{\mathrm{p}}}
\newcommand{\hatex}{\hat{\textbf{e}}_x}
\newcommand{\hatey}{\hat{\textbf{e}}_y}
\newcommand{\hatez}{\hat{\textbf{e}}_z}
\newcommand{\Phis}{\Phi_{\textrm{s}}}
\newcommand{\bfOmegap}{{\bm \Omega}_{\rm p}}
\newcommand{\bfOmega}{{\bm \Omega}}

\newcommand{\vceta}{v_{{\rm c}\eta}}
\newcommand{\vcxi}{v_{{\rm c}\xi}}
\newcommand{\vseta}{v_{{\rm s}\eta}}
\newcommand{\vsxi}{v_{{\rm s}\xi}}
\newcommand{\vsx}{v_{{\rm s}x}}
\newcommand{\vcx}{v_{{\rm c}x}}
\newcommand{\vsy}{v_{{\rm s}y}}
\newcommand{\vcy}{v_{{\rm c}y}}
\newcommand{\vco}{v_{{\rm c}0}}
\newcommand{\vcxo}{v_{{\rm c}x0}}
\newcommand{\vcyo}{v_{{\rm c}y0}}
\newcommand{\rhoc}{\rho_{\rm c}}
\newcommand{\rhos}{\rho_{\rm s}}

\newcommand{\xs}{x_{\mathrm{s}}}

\newcommand*\widefbox[1]{\fbox{\hspace{2em}#1\hspace{2em}}}

\def\mv{\mathbf{v}}

\def\d{\mathrm{d}}

\def\pa{\partial}

\def\aap{A\&A}\def\apj{ApJ}\def\mnras{MNRAS}\def\apjs{ApJS}\def\apjl{ApJ}

\title[Origin and Properties of the Wiggle Instability] {The Physical Origin and the Properties of Arm Spurs/Feathers in Local Simulations of the Wiggle Instability}
\author[Mandowara, Sormani, Sobacchi \& Klessen]{Yash Mandowara$^{1}$, Mattia C. Sormani$^{1}$, Emanuele Sobacchi$^{2}$ and Ralf S. Klessen$^{1,3}$\\
$^1$ Universit\"{a}t Heidelberg, Zentrum f\"{u}r Astronomie, Institut f\"{u}r theoretische Astrophysik, Albert-Ueberle-Str. 2, 69120 Heidelberg, Germany \\
$^2$ Department of Astronomy and Columbia Astrophysics Laboratory, Columbia University, 550 West 120th Street New York, NY 10027, USA \\
$^3$ Universit\"{a}t Heidelberg, Interdiszipli\"{a}res Zentrum f\"{u}r Wissenschaftliches Rechnen, Im Neuenheimer Feld 205, 69120 Heidelberg, Germany
}

\begin{document}

\date{}
\maketitle

\begin{abstract}
Gaseous substructures such as feathers and spurs dot the landscape of spiral arms in disc galaxies. One of the candidates to explain their formation is the wiggle instability of galactic spiral shocks. We study the wiggle instability using local 2D hydrodynamical isothermal non-self gravitating simulations. We find that: (1) Simulations agree with analytic linear stability analysis only under stringent conditions. They display surprisingly strong non-linear coupling between the different modes, even for small mode amplitudes ($\sim 1\%$). (2) We demonstrate that the wiggle instability originates from a combination of two physically distinct mechanisms: the first is the Kelvin-Helmholtz instability, and the second is the amplification of infinitesimal perturbations from repeated shock passages. These two mechanisms can operate simultaneously, and which mechanism dominates depends on the underlying parameters. (3) We explore the parameter space and study the properties of spurs/feathers generated by the wiggle instability. The wiggle instability is highly sensitive to the underlying parameters. The feather separation decreases, and the growth rate increases, with decreasing sound speed, increasing potential strength and decreasing interarm distance. (4) We compare our simulations with a sample of 20 galaxies in the HST Archival Survey of Spiral Arm Substructure of La Vigne et al. and find that the wiggle instability is able to reproduce the typical range of feather separations seen in observations. It remains unclear how the wiggle instability relates to competing mechanisms for spur/feather formation such as the magneto-jeans instability and the stochastic accumulation of gas due to correlated supernova feedback.
\end{abstract}
  
\section{Introduction}

Gaseous spiral arms in galaxies often exhibit substructure such as spurs and feathers that extend from the arms into the interarm region \citep[e.g.][]{LaVigne2006,Leroy2017,Sch2017,Kreckel2018,Elmegreen2019}. The formation of these substructures is not fully understood. Various mechanisms have been proposed, including gravitational instability/amplification of perturbations in the arm crest \citep{Elmegreen1979,Cowie1981,Bal1985,Balbus1988,Elmegreen1994,Wada2008,Henshaw2020}, magneto-gravitational instabilities \citep{Kim2002,Kim2006,Lee2012,Lee2014}, and clustering of gas due to correlated supernova feedback \citep{Kim2020}.

One possible mechanism that does not rely on gas self-gravity and magnetic fields is the wiggle instability \citep{Wada04}. The wiggle instability is a purely hydrodynamic instability that causes spiral shock fronts to fragment and break into regularly spaced ``plumes''. The morphology of these plumes resembles that of spurs/feathers observed along the spiral arms of disc galaxies. However, the physical origin of the wiggle instability, and how the underlying parameters influence the properties of the plumes, are not completely understood.

\subsection{A brief history of the stability of galactic spiral shocks} \label{sec:history}

The theoretical study of galactic spiral shocks goes back several decades. \citet{Fugi1966} and \citet{Roberts1969} \citep[see also][]{LinShu1964,Shu1973} first demonstrated that the interstellar gas can develop strong spiral shocks in response to an externally imposed spiral gravitational potential, even for modest amplitudes of the background spiral potential. These works had a strong impact on the star formation community since it was realised that these spiral shocks could trigger the gravitational collapse of molecular clouds, leading to the formation of new stars. These authors derived stationary (i.e., steady-state) solutions for the gas flow, but did not address the question of the stability of these solutions.

Several works subsequently investigated the stability of the spiral shocks, aiming to understand how this might affect the star formation process and the formation of substructures along the spiral arms. Early theoretical analyses generally concluded that the shocks are stable \citep{Nelson1977,Bal1985,Balbus1988,Dwarkadas1996}. It therefore came as a surprise when \cite{Wada04}, using global 2D non-self gravitating isothermal simulations of gas flow in a spiral potential, found that the spiral shocks are hydrodynamically unstable. The spiral shock fronts in their simulations develop wiggles and clumps, and these authors dubbed this phenomenon the ``wiggle instability''. The instability has subsequently been observed in numerous other numerical simulations \citep[e.g.][]{Wada2008,Kim2012,Kim2014,Sormani2015,Fragkoudi2017}.

The physical origin of the wiggle instability observed by \cite{Wada04} is debated. Three main hypotheses have been put forward: 
\begin{enumerate}[leftmargin=*]
\item The wiggle instability is a manifestation of the familiar Kelvin-Helmholtz instability (KHI), occurring in the post-shock region where the gas shear is very high \citep{Wada04}.
\item The wiggle instability is caused by the amplification of perturbations through multiple shock passages \citep{Kim32014,Kim2015,Sormani2017}. 
\item The wiggle instability is a numerical artefact, e.g.\ caused by the discretisation of the fluid equations in hydrodynamical schemes \citep{Hanawa12}.
\end{enumerate}

\subsection{Aims of this paper}

This paper aims to elucidate the physical origin of the wiggle instability using local idealised 2D hydrodynamical simulations and to study the properties of the generated substructure as a function of the underlying parameters. In particular, we address the following questions:
\begin{itemize}[leftmargin=*]
\item Which of the three hypotheses for the origin of the wiggle instability mentioned in Sect.~\ref{sec:history} is correct?
\item If the instability is real (as we will argue in this paper), why did the early theoretical studies \citep{Nelson1977,Bal1985,Balbus1988,Dwarkadas1996} conclude that the shocks are stable?
\item How do the properties of the wiggle instability (feather separation, growth rate) depend on the underlying parameters (gas sound speed, strength of the spiral arm potential, interarm separation, background shear)?
\item Can the wiggle instability reproduce the observed properties of spurs/feathers in real spiral galaxies?
\end{itemize}

As we will see, our answer to the first of these questions is that the wiggle instability, i.e.\ the phenomenon of unstable behaviour of the spiral shocks observed in unmagnetised non-self-gravitating simulations, originates from a combination of two physically distinct mechanisms: the first is the Kelvin-Helmholtz instability (item i in Sect.~\ref{sec:history}), and the second is the amplification of infinitesimal perturbations from repeated shock passages (item ii in Sect.~\ref{sec:history}). These two mechanisms can operate simultaneously, and which mechanism dominates depends on the parameter of the system under consideration.

This paper is structured as follows. In Section \ref{sec:equations}, we present the formulation of the problem. In Section \ref{sec:setup} we describe our numerical setup and our methodology for the shock front analysis. In Section \ref{sec:linear} we compare in detail three example simulations with predictions from the linear stability analysis of \citet{Sormani2017} by performing a Fourier decomposition of the unstable shock front as a function of time. In Section \ref{sec:KH} we demonstrate that boundary conditions are critical for the development of the wiggle instability, and we prove that different physical mechanisms are responsible for the wiggle instability in different parameter regimes. In Section \ref{sec:scan} we perform a parameter space study. In Section \ref{sec:discussion} we compare our results with HST observations of spurs/feathers in disc galaxies and we discuss the relation of the wiggle instability with other proposed mechanisms of spurs/feather formation.  We sum up in Section \ref{sec:conclusion}.

\section{Basic equations} \label{sec:equations}

Our goal is to study the wiggle instability in the simplest possible setup, to understand its physical origin in the clearest possible way. Following \citet{Roberts1969} we approximate the equations of hydrodynamics in a local Cartesian patch that is corotating with a segment of a spiral arm located at galactocentric radius $R=R_0$. We briefly summarise the setup here and offer a detailed derivation of the equations in Appendix \ref{sec:derivation}.

The gas is assumed to flow in an externally imposed gravitational potential that is the sum of an axisymmetric component with circular angular velocity $\Omega(R)$ plus a spiral perturbation $\Phis$ that rigidly rotates with pattern speed $\Omegap$. We assume an isothermal equation of state $P=\cs^2 \rho$, where $P$ is the pressure, $\rho$ is the surface density and $\cs$ is the (constant) sound speed, and that the gas is two-dimensional, non-self-gravitating and unmagnetised.

We call $(x,y)$ the local Cartesian coordinates in a frame that is sliding along the arm segment with a speed equal to the local circular velocity, where $x$ is the coordinate perpendicular to the arm and $y$ the coordinate parallel to the arm. The equations of motion in this frame approximated under the assumption of small pitch angle ($\sin(i)\ll 1$) are (see Appendix \ref{sec:derivation}):
\begin{align} \label{eq:2e1}
\pa_t\mv + ( \mv\cdot\nabla)\mv &= -\frac{\nabla P}{\rho} - \nabla\Phis -2\bfom_0\times\mv + q\Omega_0 v_x \hatey + F \left( 1 - \frac{q}{2} \right) \hatey\\
\pa_t\rho+\nabla\cdot(\rho\mv) &= 0, \label{eq:2e2}
\end{align}
where $\bfv = v_x \hatex + v_y\hatey$ is the velocity in the $(x,y)$ frame (in other words, $v_x$ is the velocity perpendicular to the arm in the frame corotating with the spiral arms, while $v_y$ is the difference between the velocity parallel to the arm and the local circular velocity), $\Omega_0 = \Omega(R_0)$, $\bfom_0=\Omega_0 \hatez$, 
\begin{equation}
q = - \left( \frac{\d \log \Omega}{\d \log R} \right)_{R=R_0},
\end{equation}
is the shear parameter calculated at $R=R_0$\,,
\begin{equation}
F = 2\Omega_0 \vco \sin(i)\,,
\end{equation}
is a constant, and $\vco=(\Omega_0-\Omegap)R_0$ is the circular velocity at $R_0$ in the frame corotating with the spiral arms.

We assume that in our local frame the spiral perturbation to the potential has the form:
\begin{equation}
\Phis = \Phi_0 \cos\bigg(\frac{2\pi x}{L_x}\bigg), \label{eq:51}
\end{equation}
where $\Phi_0$ is constant and $L_x \ll R_0$ is the size of our local patch in the $x$ direction, which is equal to the separation between two consecutive spiral arms at $R=R_0$ (see Eq.~\ref{eq:L2}). Note that, since we assume that the spiral potential only depends on $x$, our system is transitionally invariant in the $y$ direction.

Equations \eqref{eq:2e1} and \eqref{eq:2e2} are identical to those of a simple 2D fluid that is subject to the following forces:
\begin{enumerate}[leftmargin=*,label=\arabic*.]
\item The pressure $-\nabla P /\rho$.
\item The external spiral potential $-\nabla \Phi$.
\item The Coriolis force $-2\bfOmega_0\times\mv$.
\item A constant force $F (1-q/2)\hatey$.
\item The ``shear'' force $q\Omega_0 v_x \hatey$.
\end{enumerate}

\subsection{Parameters}\label{sec:params}

The problem posed by Equations \eqref{eq:2e1} and \eqref{eq:2e2} is completely specified by the six parameters $\{\cs,\Phi_0,L_x,q,\Omega_0,F\}$. From these, we can define four dimensionless parameters and two scaling constants. Without loss of generality, we choose to use $\Omega_0$ and $F$ as scaling constants. We rescale the others as indicated in Table \ref{tab:final_params} to obtain four independent dimensionless parameters $\{\tilde{c}_{\rm s}, \tilde{\Phi}_0, \tilde{L}_x, \tilde{q}\}$. In this paper, we explore how the properties of the wiggle instability depend on these four. For simplicity of notation, we drop the `tilde' superscript hereafter and always refer to the dimensionless parameters unless otherwise specified. The range of values explored in this work is indicated in Table \ref{tab:final_params}. The table indicates both the dimensionless values and the physical values calculated assuming typical galactic values for the two scaling constants. The physical value of $F$ indicated in the table corresponds to a typical circular velocity in the frame corotating with the spiral arms of $\vco=100\kms$ for a pitch angle $\sin(i)=0.1$. The explored values of $q$ include the cases of solid body rotation ($q=0$), a flat rotation curve ($q=1$), and a Keplerian rotation curve ($q=1.5$).

Using the typical values of the scaling constants indicated in Table \ref{tab:final_params}, one unit of dimensionless time corresponds to $1/\Omega_0 = 48.9 \Myr$ of physical time, one unit of dimensionless length corresponds to $F/\Omega_0^2 = 1\kpc$, and one unit of dimensionless velocity corresponds to $F/\Omega_0=20\kms$.

\begin{table*}
\begin{center}
\caption{Parameters of the problem.}
\label{tab:final_params}
\begin{tabular}{c|c|c|c|c}
\textbf{Parameter} & \textbf{Brief description} & \textbf{Physical values} & \textbf{Dimensionless formulation} & \textbf{Dimensionless values}\\
\hline
\rowcolor{gray!00} $\cs$ & isothermal sound speed & 4 - 14 $\kms$ & \parbox{3cm}{\[\tilde{c}_{\rm s} = \frac{\cs}{F/\Omega_0}\]} & 0.2 - 0.7\\
\rowcolor{gray!20} $\Phi_0$ & strength of the spiral potential & 10 - 100 $(\kms)^2$ & \parbox{3cm}{\[\tilde{\Phi}_0 = \frac{\Phi_0}{(F/\Omega_0)^2}\]} & 0.025 - 0.25 \\
\rowcolor{gray!00} $L_x$ & spiral arm separation & 0.4 - 2.0 kpc & \parbox{3cm}{\[\tilde{L}_x = \frac{L_x}{(F/\Omega_0^2)}\]} & 0.4 - 2.0\\
\rowcolor{gray!20} $q$ & shear factor & 0 - 1.5 & \parbox{3cm}{\[\tilde{q}=q\]} & 0 - 1.5\\
\rowcolor{gray!00} $F$ & background Coriolis force $\perp$ to arm & 400\;$(\kms)^2$ kpc$^{-1}$ & \parbox{3cm}{\[ \tilde{F}= \frac{F}{F}\]} & 1 \\
\rowcolor{gray!20} $\Omega_0$ & local rotation velocity & 20 $\kms$ kpc$^{-1}$ & \parbox{3cm}{\[\tilde{\Omega}_0=\frac{\Omega_0}{\Omega_0}\]} & 1\\
\end{tabular}
\end{center}
\end{table*}

\subsection{steady-state solution}\label{sec:sss}

Equations \eqref{eq:2e1} and \eqref{eq:2e2} admit steady-state solutions that are periodic in the $x$ coordinate and do not depend on the $y$ coordinate. As noted by \citet{Roberts1969} (see also \citealt{Shu1973}), these steady-state solutions must contain a shock if $\Phi_0$ is above a critical value (this critical value depends on the values of the other parameters). These shocked steady-state solutions constitute the initial conditions for the simulations described below. This paper aims to see under which conditions these solutions are prone to the development of the wiggle instability and to study the properties of the substructures generated by the instability.

Assuming steady-state, Equations \eqref{eq:2e1} and \eqref{eq:2e2} reduce to the following system of ordinary differential equations:
\begin{align}
\frac{\d v_{x}}{\d x} & = \left[ 2\Omega_0 v_y - \frac{\d \Phis}{\d x} \right] \left( v_x-\frac{\cs^2}{v_x}\right)^{-1}
,\label{eq:2A1}\\
\frac{\d v_{y}}{\d x}&=(q-2) \left(\Omega_0 v_x + \frac{F}{2}\right) \,.\label{eq:2A2}
\end{align}
The density is given by $\rho(x)= C/v_x$, where $C$ is an arbitrary constant which without loss of generality we set to unity. Note that the solution for $\Phis=0$ is $v_x=-F/(2\Omega_0)=\vco\sin(i)$, $v_y=0$ ($v_x=1/2$, $v_y=0$ in dimensionless units). This solution corresponds to the fact that in absence of the spiral perturbation to the potential, the component of the circular motion perpendicular to the spiral arm is simply $\vco\sin(i)$. Note that this solution is subsonic if $\cs>0.5$ and supersonic if $\cs<0.5$ (in dimensionless units).

The system of equations \eqref{eq:2A1} and \eqref{eq:2A2} are solved numerically using the shooting method. The numerical procedure is described in more detail in Appendix \ref{sec:steady}. 

\section{Methodology} \label{sec:setup}
\subsection{Simulation setup} \label{sec:bc}

We solve the equations of hydrodynamics with the public grid code {\sc PLUTO} version 4.3 \citep{pluto}. We use a two-dimensional static Cartesian grid with uniform spacing $\Delta x=\Delta y=1.25\times 10^{-3}$ in dimensionless units (corresponding to $1.25\pc$ in physical units). The size of the computational box is $L_x \times L_y$. $L_x$ is varied within the range indicated in Table \ref{tab:final_params}. We adopt $L_y=2$ for most of the simulations reported in this paper, but we experiment with different values in Section \ref{sec:examples}. These box sizes correspond to $N_x = 320$-$1600$ grid points in the $x$ direction (depending on the value of $L_x$) and $N_y=1600$ grid points in the $y$ direction. We use the following parameters within the {\sc PLUTO} code: RK2 time-stepping, no dimensional splitting, isothermal equation of state, Roe Riemann Solver, and the default flux limiter. The time-step is determined according to the Courant-Friedrichs-Lewy (CFL) criterion, with a CFL number of 0.4.

The initial conditions are provided by the steady-states described in Section~\ref{sec:sss}. We let the system evolve till $t_f=30.0$ in dimensionless units, corresponding to $t_f\simeq1.5\Gyr$ in physical units. We introduce some random seed noise in the initial conditions to accelerate the onset of instability and save significant computational time. The instability would develop at a later time even without this initial noise, and we have tested that the properties of the induced substructure (morphology and feather separation) are unaffected by the introduction of the noise. The way in which noise is introduced is described in detail in Appendix~\ref{sec:noise}.

We use two types of boundary conditions in this work. The first is periodic boundary conditions in both the $x$ and $y$ directions. This type of boundary condition is the most appropriate for galactic spiral shocks because it takes into account the fact that the material leaving one spiral arm will later pass through the next spiral arm. The second type of boundary condition is inflow-outflow, also known as D'yakov-Kontoroich (DK) boundary conditions after the classic shock front stability analysis of \citet{Dy1954} and \citet{Kont1958}. With DK boundary conditions, we assume a constant injection of gas at the $x=0$ boundary at the rate given by the steady-state solution (Section~\ref{sec:sss}), while gas can freely escape at the $x=L_x$ boundary thanks to standard outflow boundary conditions. The boundary is periodic in the $y$ direction. This second type of boundary condition is appropriate for most ``normal'' non-astrophysical circumstances, in which the pre-shock flow should be left unperturbed on account of the fact that the signal cannot travel backwards at supersonic velocity (see \S90 in \citealt{Landau}).

\subsection{Shock front analysis} \label{sec:shock}

\begin{figure*}
\centering
\includegraphics[width=\linewidth]{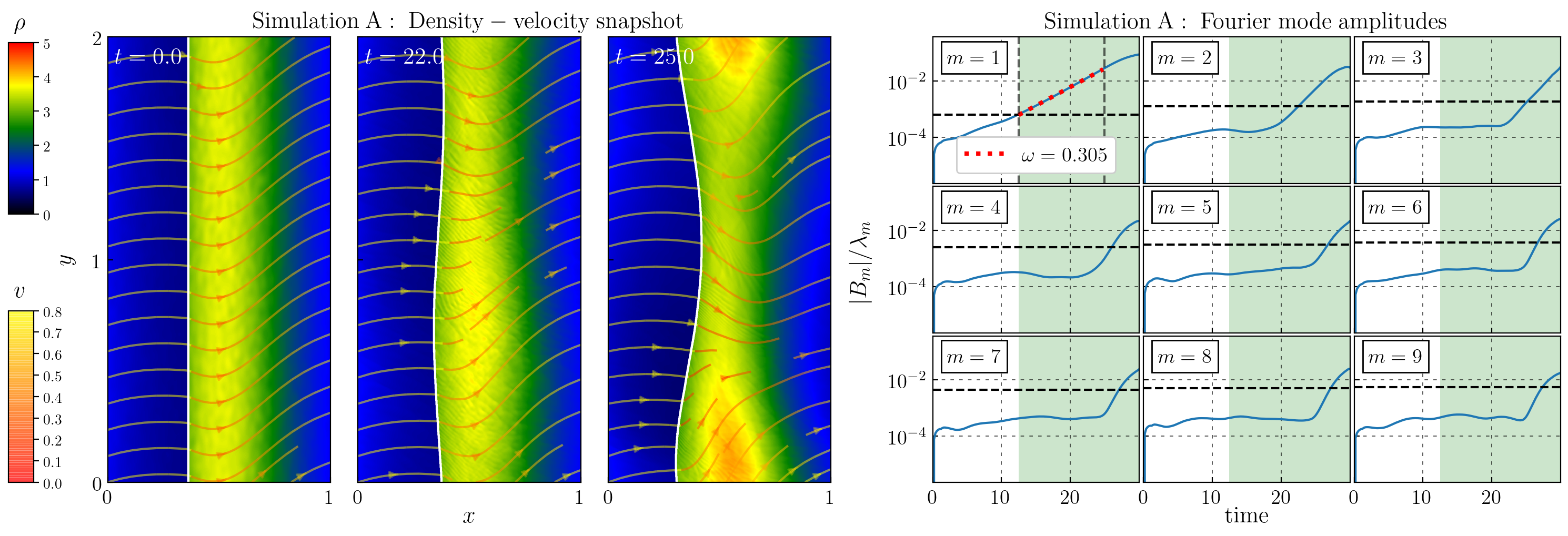}
\includegraphics[width=\linewidth]{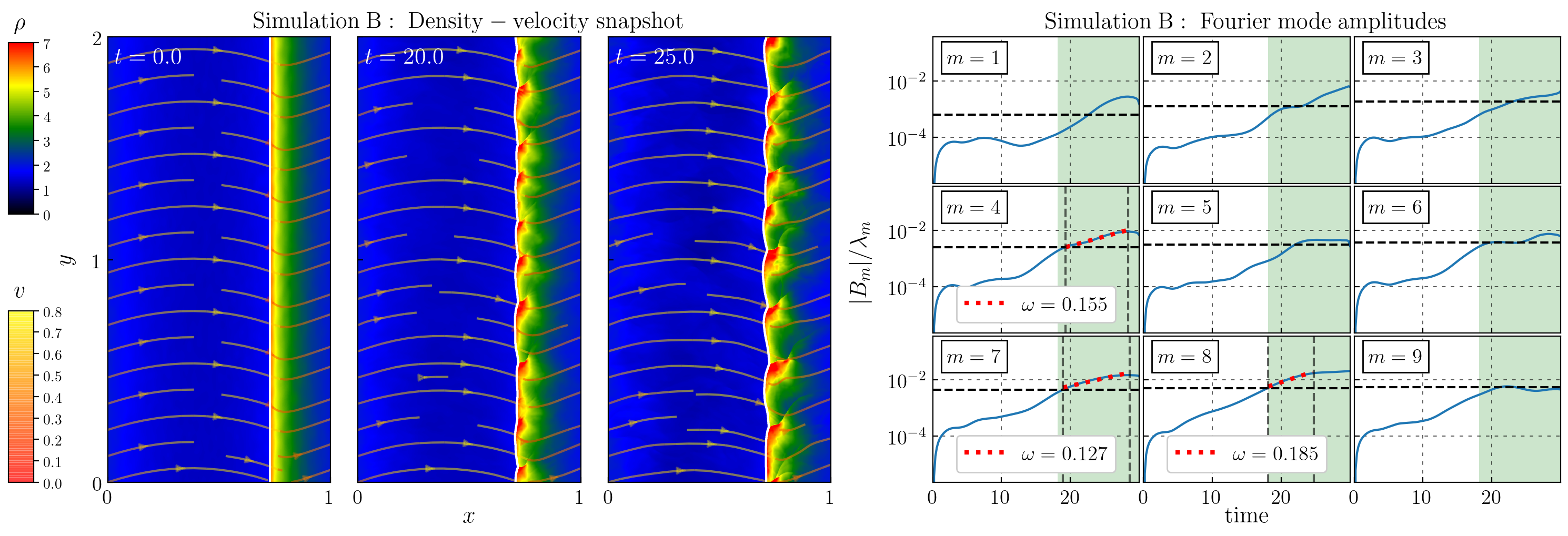}
\includegraphics[width=\linewidth]{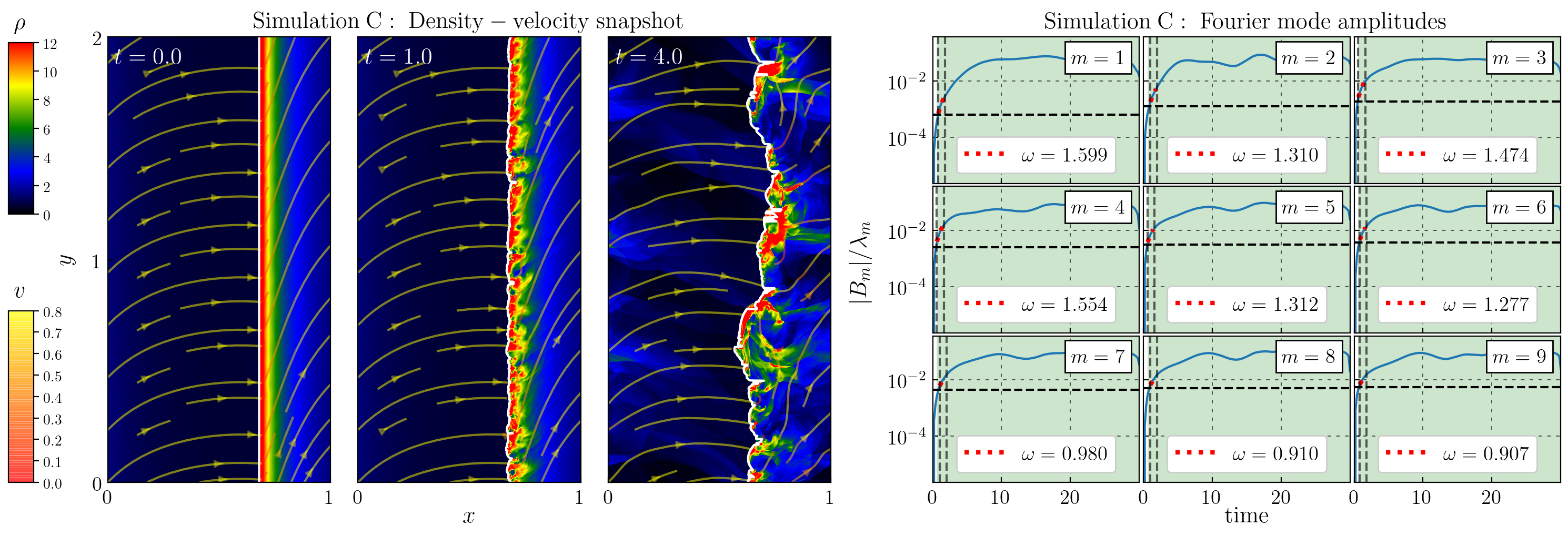}
\caption{Analysis of the three example simulations that we compare to the linear stability analysis (see Section \ref{sec:examples} and Table \ref{tab:examples}). \emph{Top}: Simulation A. \emph{Middle}: Simulation B. \emph{Bottom}: Simulation C. \emph{Left column}: time evolution of the surface density. The white line indicates the shock front detected using the method described in Section \ref{sec:shock}. The coloured lines with arrows show instantaneous streamlines. \emph{Right column}: time evolution of the amplitudes of the first nine Fourier modes, smoothed with a moving average of width $\Delta t=2.6$. The horizontal dashed black line indicates where the amplitude of each mode becomes greater than the grid resolution, $B_m>\Delta y=1.25\times10^{-3}$. The green shaded area indicates the region where instability is detectable, defined to be the region where the amplitude of at least one mode is above the black dashed line. The red dashed line indicates the best-fitting growth rate $\omega$ obtained by fitting Equation~\eqref{eq:simplefit}. The interval between the green vertical dashed lines indicates the time range used for the fit. One time unit corresponds to $\simeq48.9\Myr$ in physical units, and one spatial unit corresponds to $1$ kpc (see Section~\ref{sec:params}). Simulation A is completely dominated by the single unstable mode with $m=1$. Simulation B has multiple prominent unstable modes. In simulation C, all modes with $m<40$ are strongly unstable.}
\label{fig:examples}
\end{figure*}

\begin{figure}
\includegraphics[width=\linewidth]{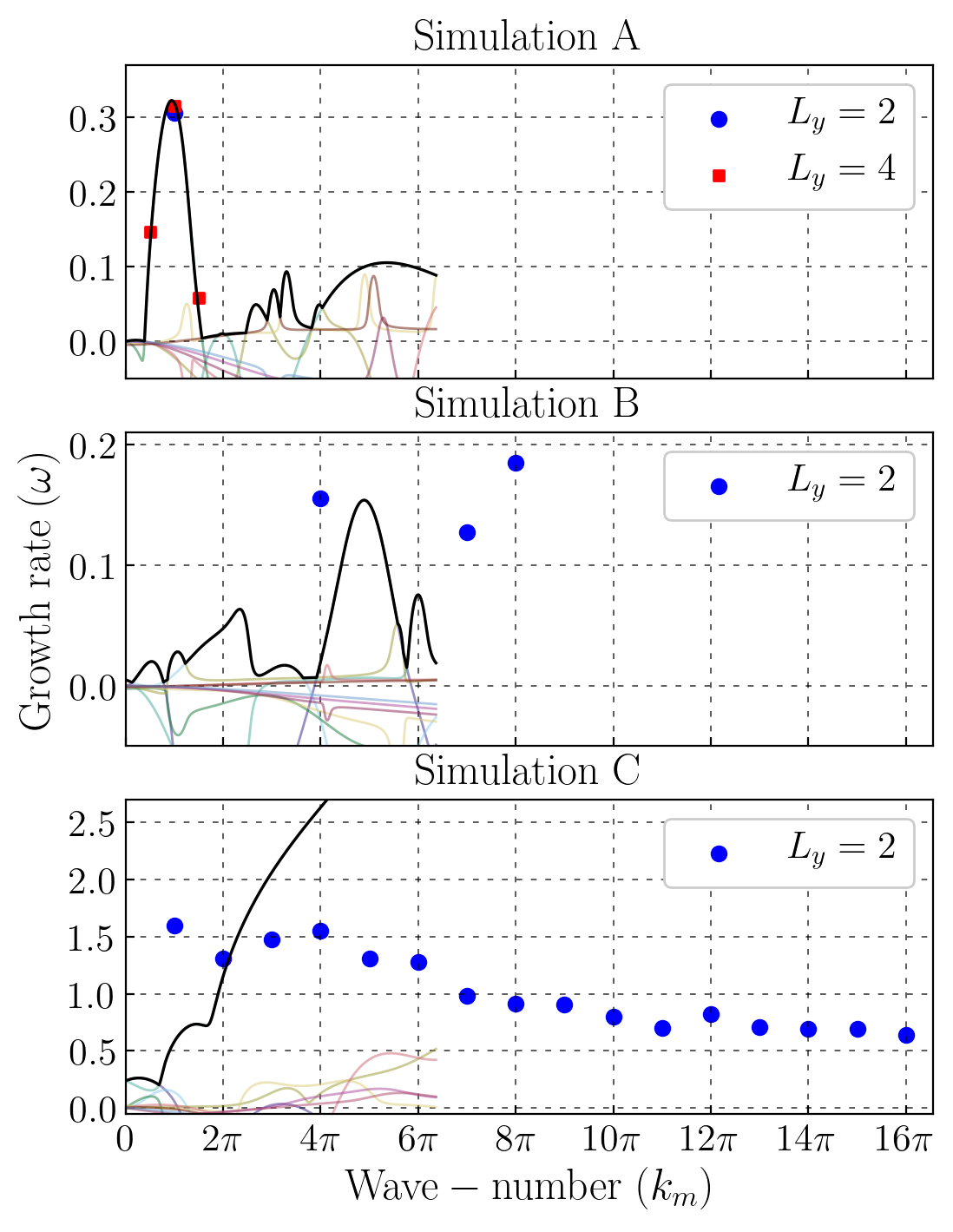}
\caption{Comparison between the growth rates measured in the simulations and predictions from the linear stability analysis. \emph{Top}: simulation A. \emph{Middle}: simulation B. \emph{Bottom}: simulation C. The solid lines indicate the dispersion relation from the linear stability analysis of \citet{Sormani2017}, which is only available for $k_m\leq20$ (see their Figures 3 and 4). The coloured lines are the individual modes, and the thick black line is the ``envelope'' of all the modes (i.e.\, it traces the mode with the highest value of $\omega$ at each wavenumber). The symbols indicate the growth rates measured in the simulations. Only discrete values of the wavenumber $k_m=2\pi m/L_y$ are present in the simulation due to the finite size $L_y$ of the box in the $y$ direction, while the linear analysis is performed in the case $L_y=\infty$. The linear analysis works well for simulation A in which the instability is caused by a single dominant mode ($m=1$ for $L_y=2$), but fails to give accurate predictions in simulations B and C due to strong coupling between unstable modes (see Sect.~\ref{sec:examples}).}
\label{fig:dispersionrelation}
\end{figure}

\begin{figure*}
\includegraphics[width=0.9\textwidth]{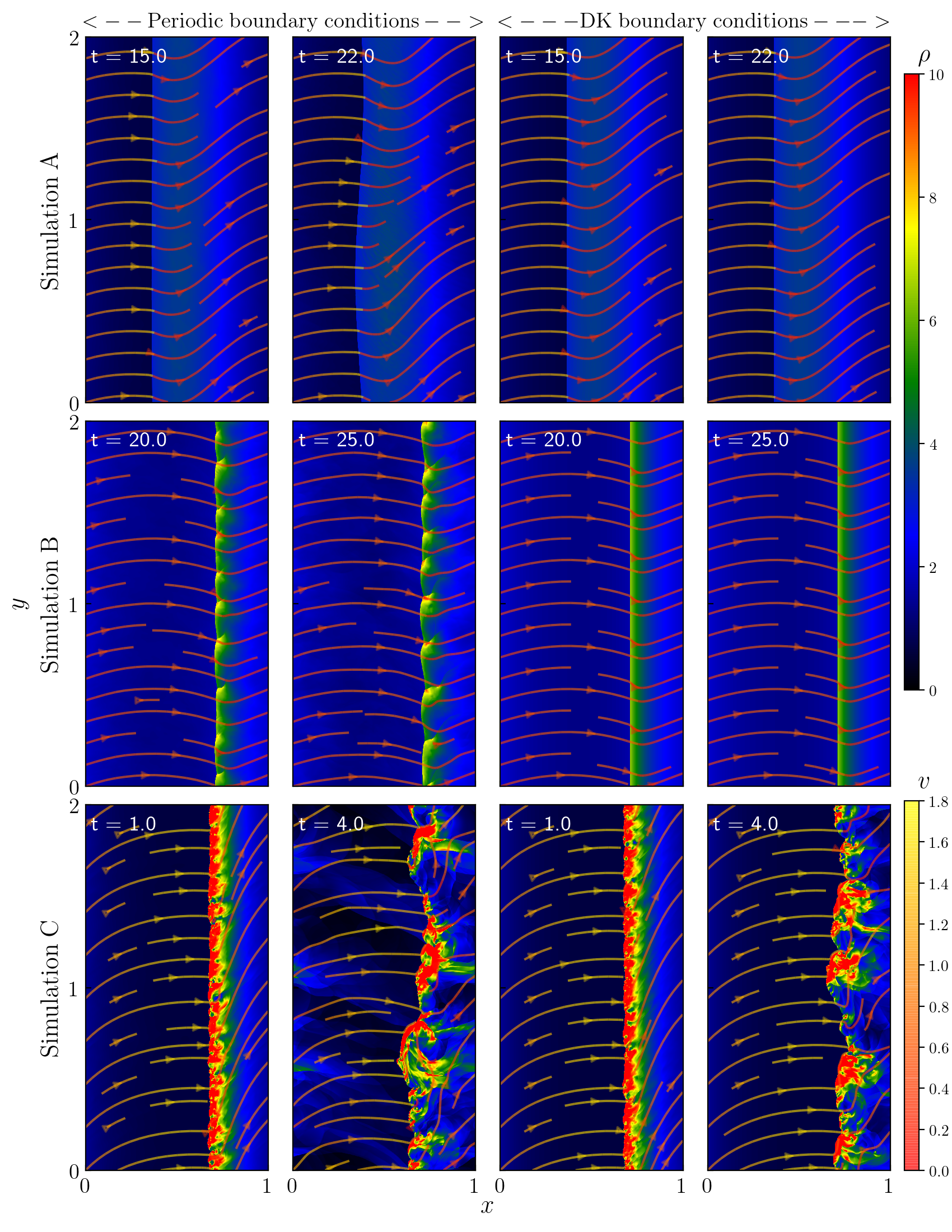}
\caption{Comparison between simulations with periodic boundary conditions (left) and inflow-outflow (DK) boundary conditions (right). \emph{Top}: simulation A. \emph{Middle}: simulation B. \emph{Bottom}: simulation C. Under periodic boundary conditions, all simulations are unstable. Under inflow-outflow (DK) boundary conditions, simulations A and B become stable, while simulation C is still unstable. This proves that the wiggle instability in simulations A \& B is purely due to the amplification of perturbations through multiple shock passages (item ii in Sect.~\ref{sec:history}), while in simulation C is at least partly due to the Kelvin-Helmholtz instability (item i in Sect.~\ref{sec:history}). See discussion in Section~\ref{sec:KH}.}
\label{fig:periodicvsdk}
\end{figure*}

\begin{figure*}
\includegraphics[width=\textwidth]{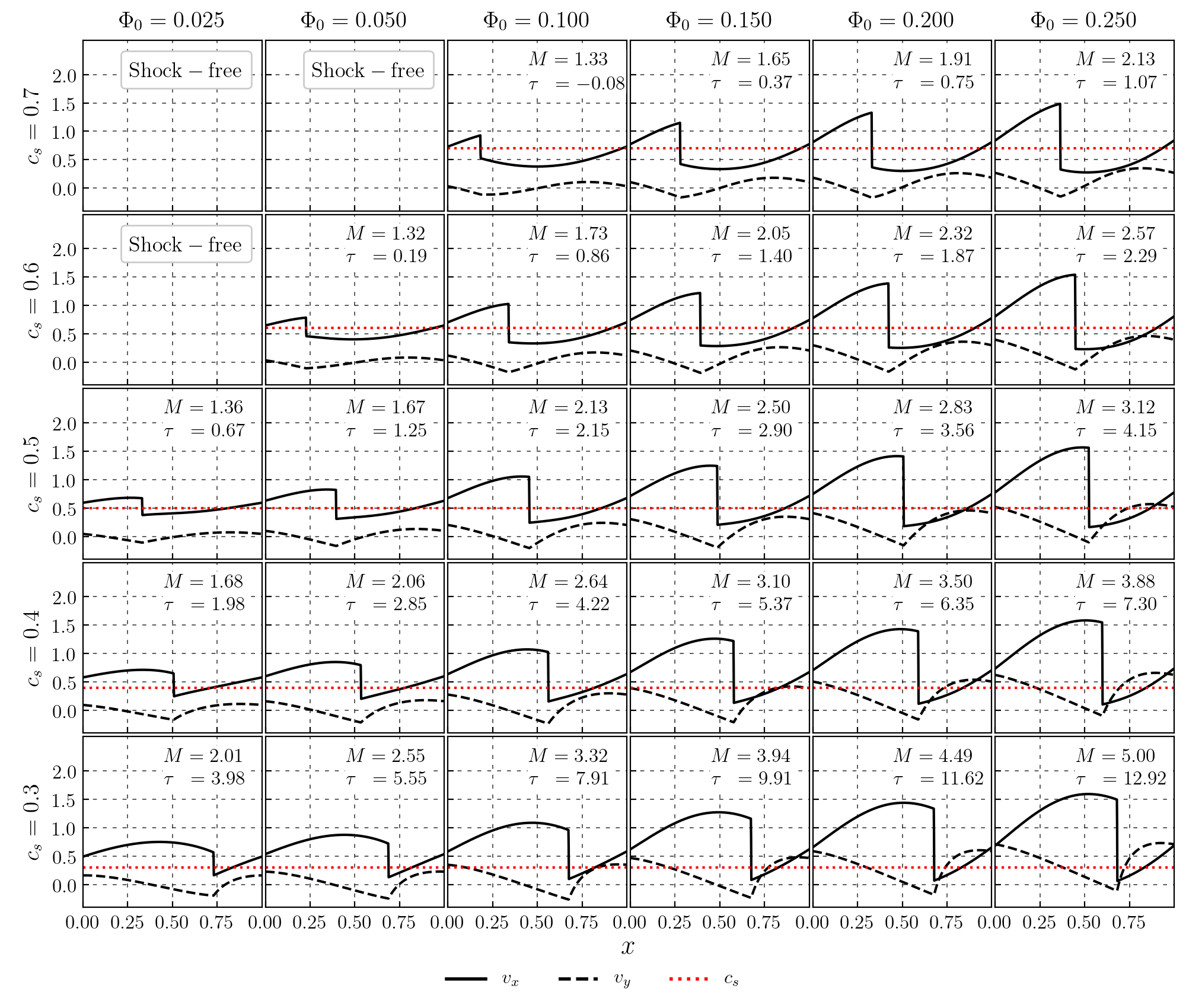}
\caption{Initial conditions for the first set of simulations (see Table \ref{tab:scan}). These initial conditions are the steady-state solutions calculated using Eqs.~\eqref{eq:2A1} and \eqref{eq:2A2}. Annotated are the Mach number $M$ and the post-shock shear $\tau$ (see caption of Table \ref{tab:scan} for definitions). The red dotted horizontal line indicates the value of the isothermal sound speed. The three panels in the upper-left corner are empty because for these values of the parameters the steady-state solutions do not contain a shock.}
\label{fig:ic1}
\end{figure*}

\begin{figure*}
\includegraphics[width=\textwidth]{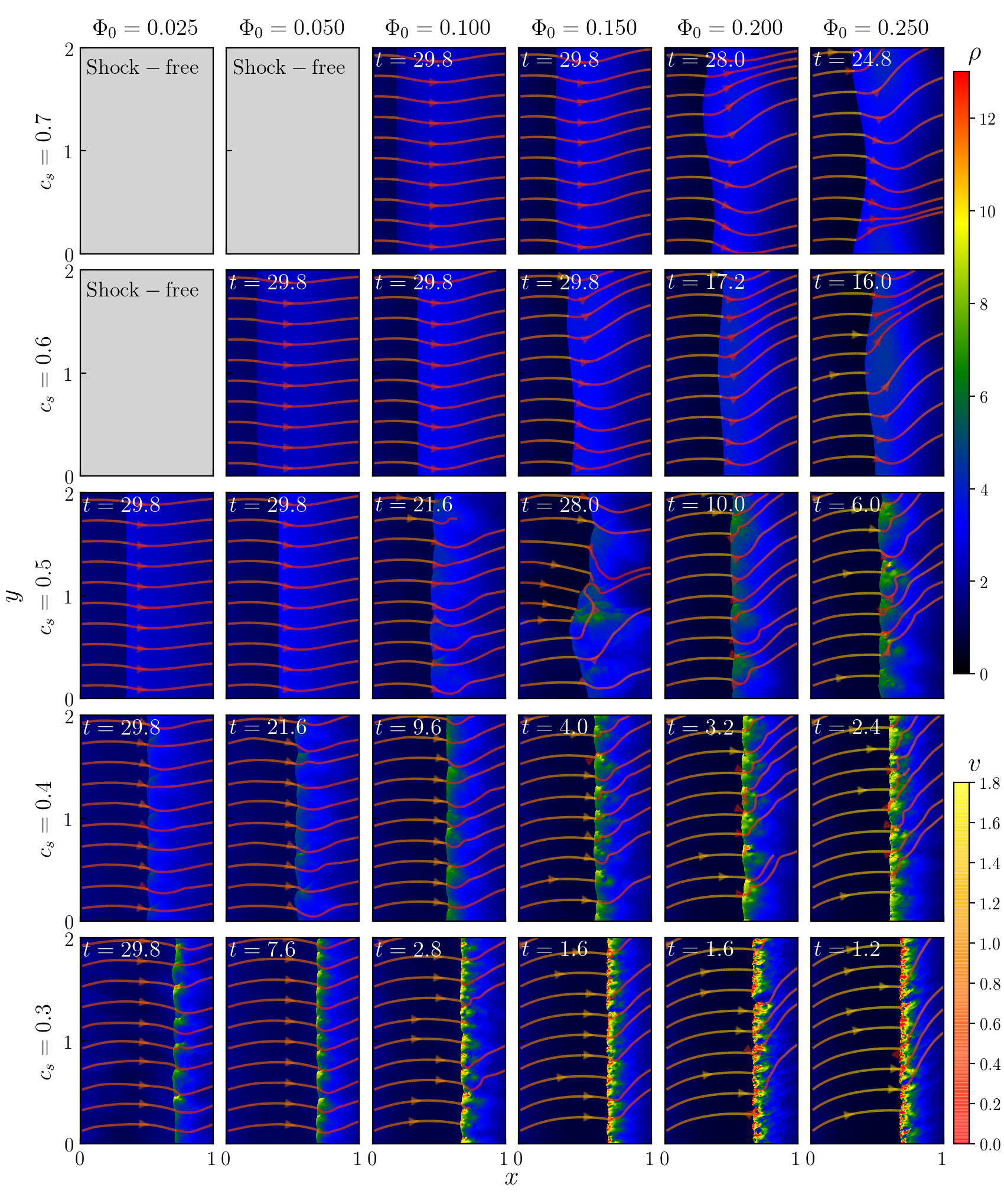}
\caption{Surface density for the first set of simulations (see Table \ref{tab:scan}), at later times with periodic boundary conditions. Instantaneous streamlines are also shown, and the colour indicates the total velocity $v=(v_x^2+v_y^2)^{1/2}$. One unit of $x$ corresponds to 1 kpc in physical units, and one unit of $t$ corresponds to $\simeq 48.9 \Myr$ (see Section~\ref{sec:params}). The three panels in the upper-left corner are empty because for these values of the parameters the steady-state solutions do not contain a shock.}
\label{fig:cpp}
\end{figure*}

\begin{figure*}
\includegraphics[width=\textwidth]{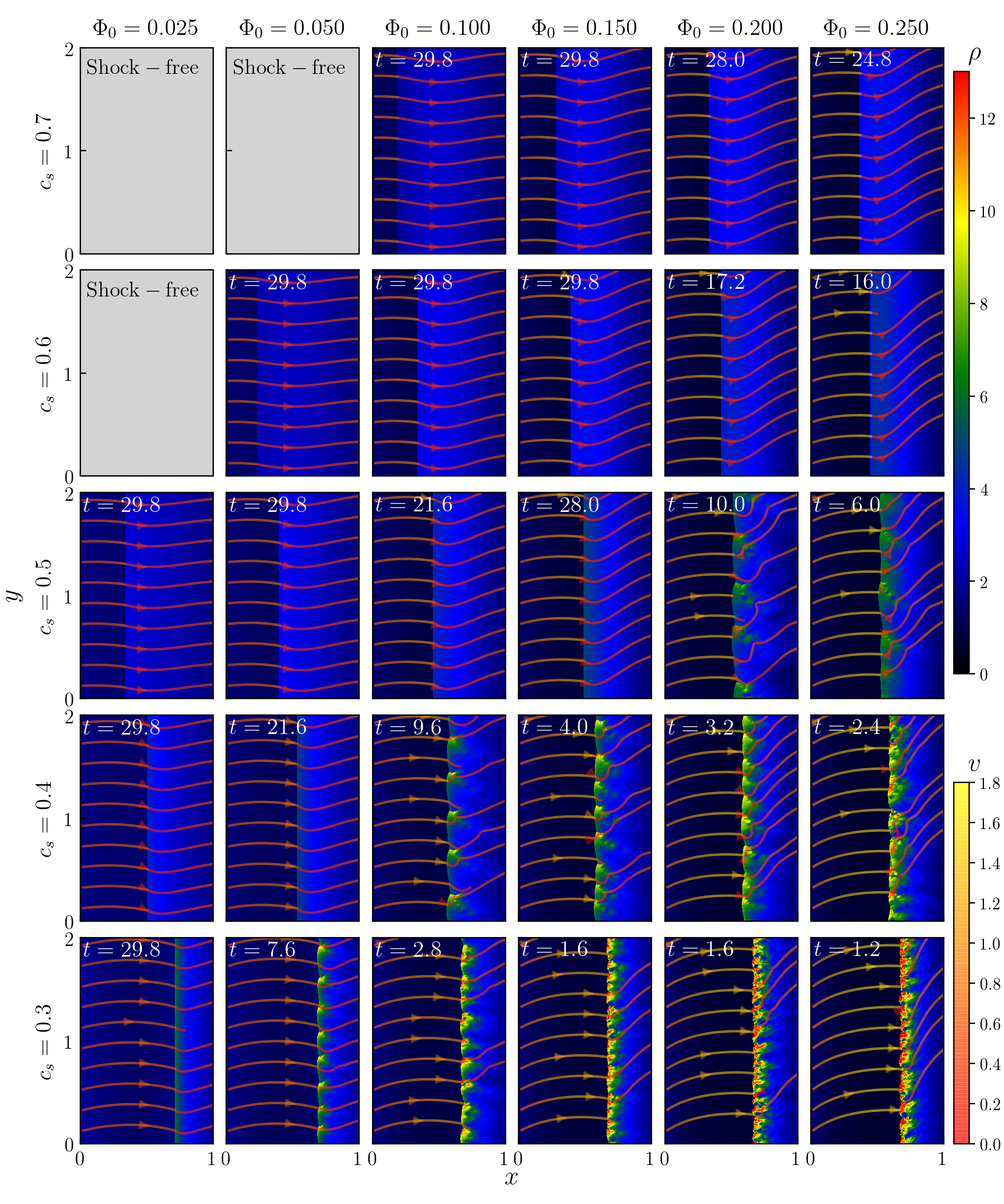}
\caption{Same as Fig.~\ref{fig:cpp}, but for inflow-outflow boundary conditions (see Sect.~\ref{sec:bc}). Only simulations in the bottom-right panels (high $\Phi_0$ and low $\cs$) are unstable, due to KHI-driven wiggle instability.}
\label{fig:cpdk}
\end{figure*}

\begin{figure*}
\includegraphics[width=\textwidth]{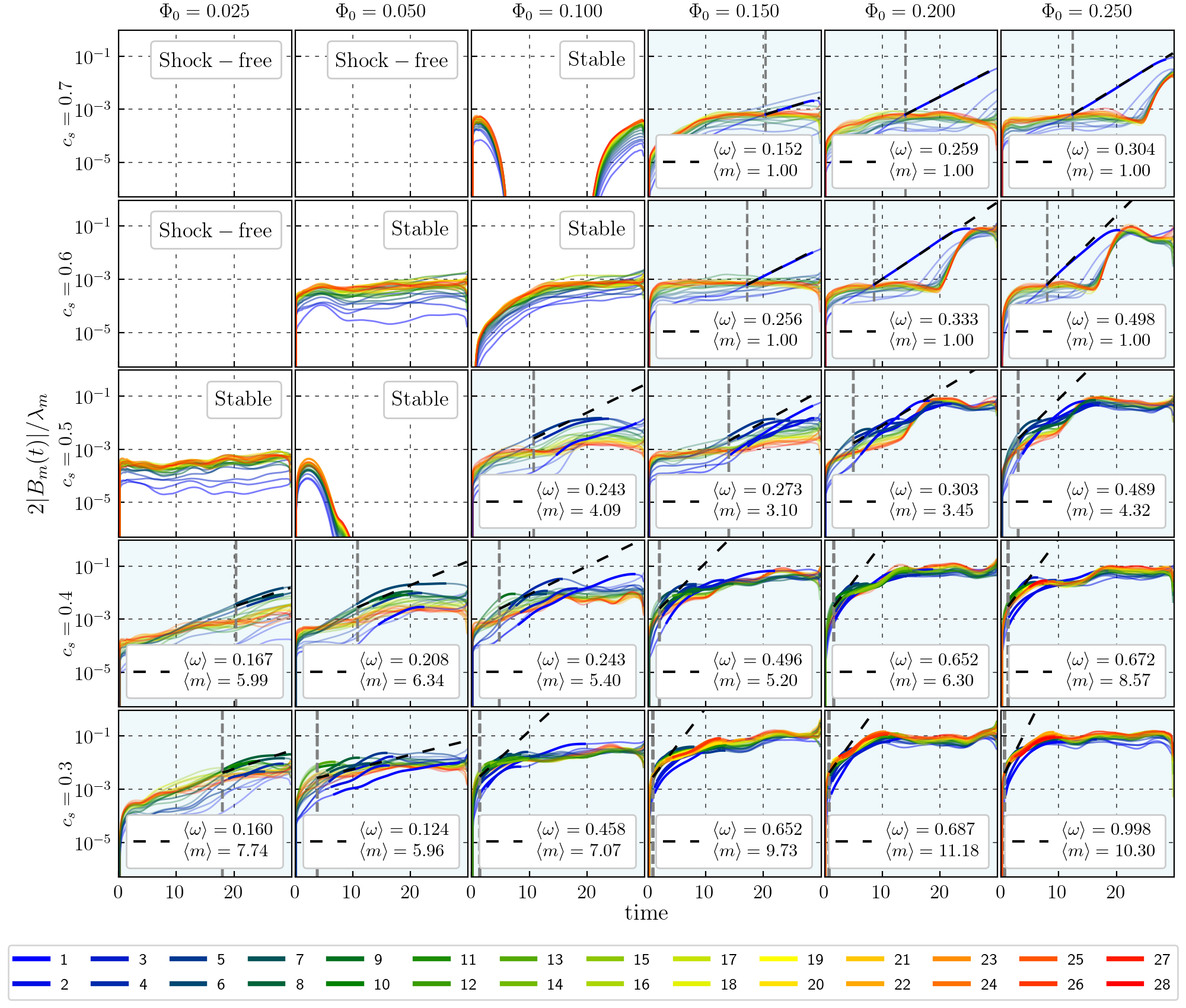}
\caption{Amplitudes of the first 28 Fourier modes as a function of time for the simulations in Figure~\ref{fig:cpp}. The average growth rate $\langle \omega \rangle$ and the average unstable wavenumber $\langle m \rangle$ is larger for larger $\Phi_0$ and smaller $\cs$. The vertical grey dashed line indicates the time when at least one amplitude becomes larger than the grid resolution. The black dashed line indicates Equation~\eqref{eq:simplefit} with the average growth rate. Blue-shaded panels indicate unstable simulations, while unshaded plots indicate simulations that do not show signs of instability within the simulation time. One time unit corresponds to $\simeq 48.9\,\Myr$ in physical units.}
\label{fig:ftcpp}
\end{figure*}

\begin{figure*}
\includegraphics[width=\textwidth]{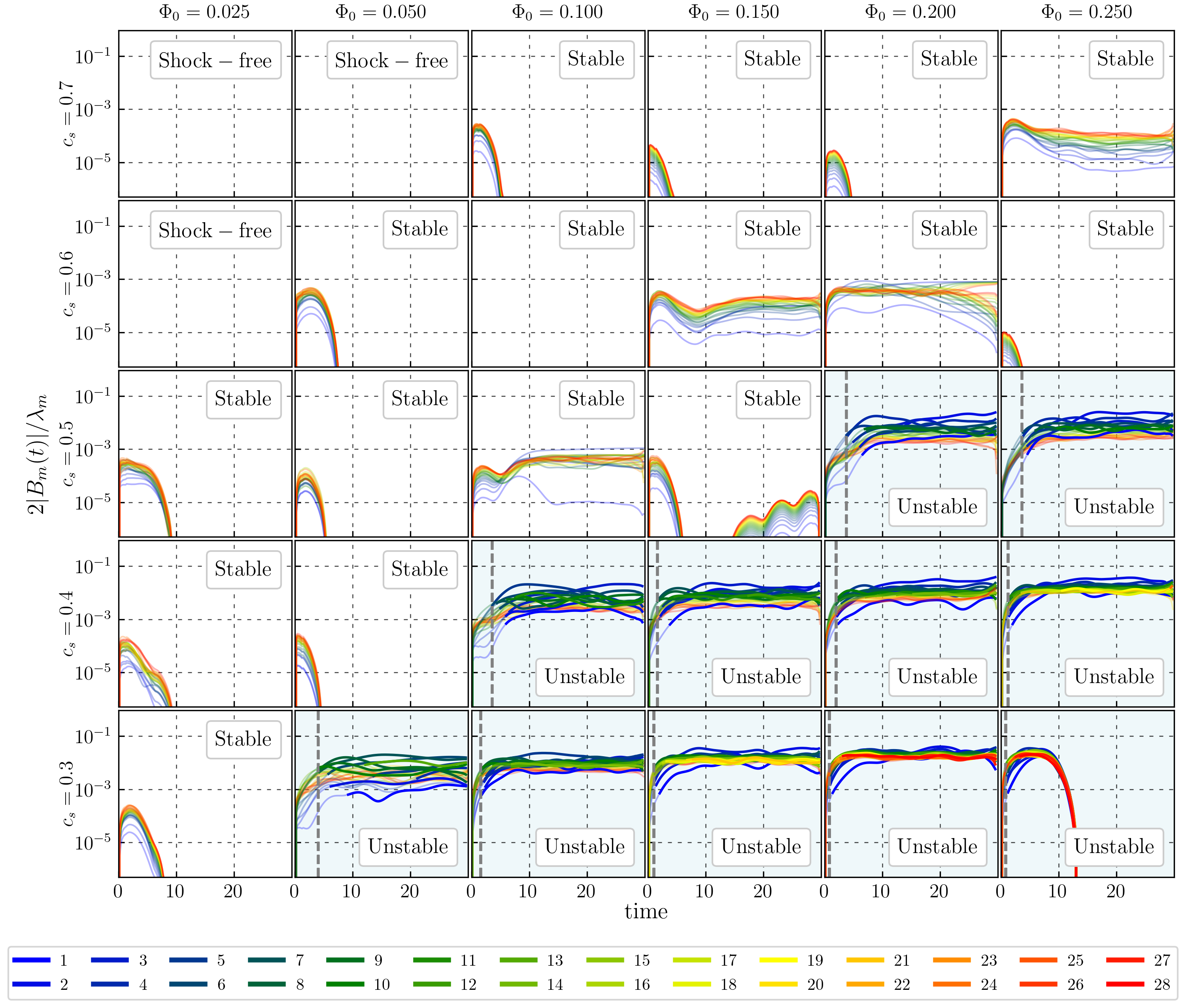}
\caption{Same as Fig.~\ref{fig:ftcpp}, but referring to the simulations in Fig.~\ref{fig:cpdk}. Note that several systems become stable when switching from periodic to inflow-outflow boundary conditions.}
\label{fig:ftcpdk}
\end{figure*}

\begin{figure*}
\includegraphics[width=0.6\textwidth]{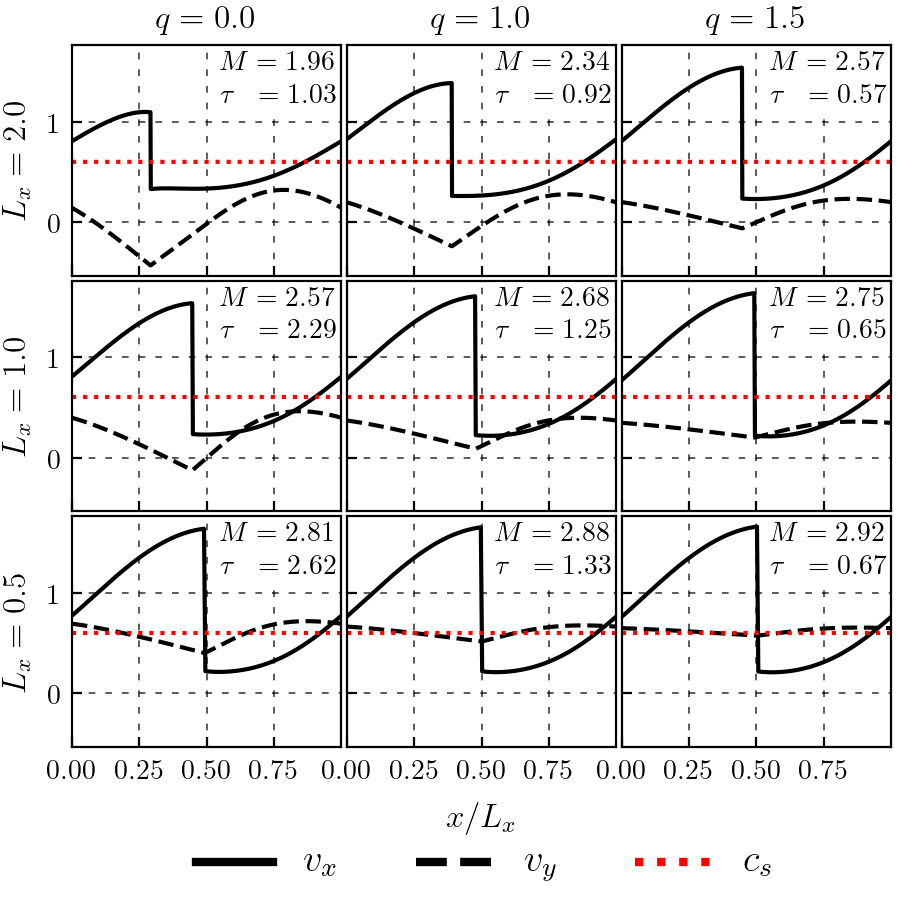}
\includegraphics[width=\textwidth]{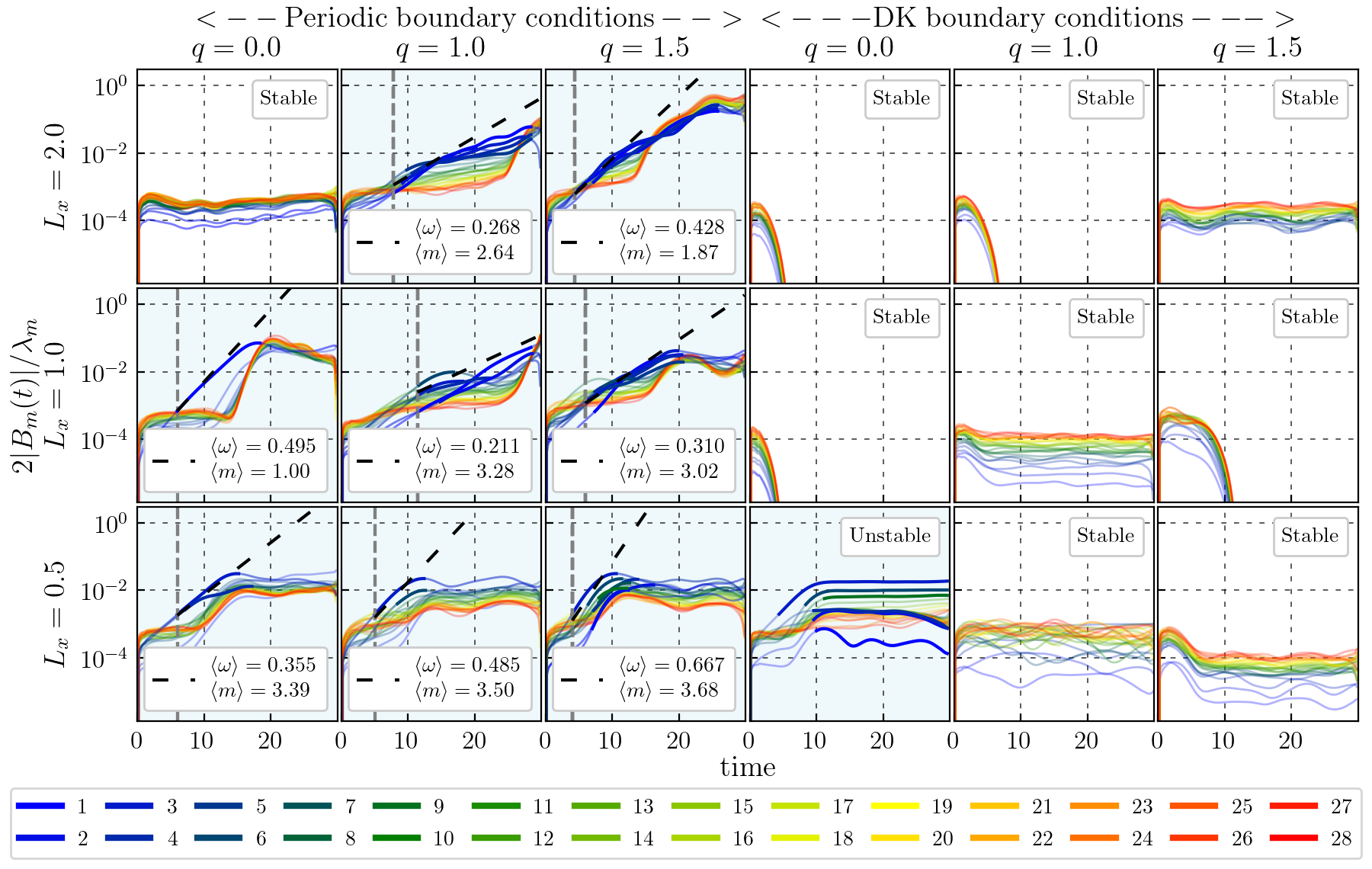}
\caption{\emph{Top}: initial conditions for the second set of simulations (see Table~\ref{tab:scan}). \emph{Bottom}: Fourier amplitudes of the first 28 modes as a function of time under periodic (left) and inflow-outflow (DK, right) boundary conditions.}
\label{fig:scanset2}
\end{figure*}

\begin{figure}
\centering
\includegraphics[width=\columnwidth]{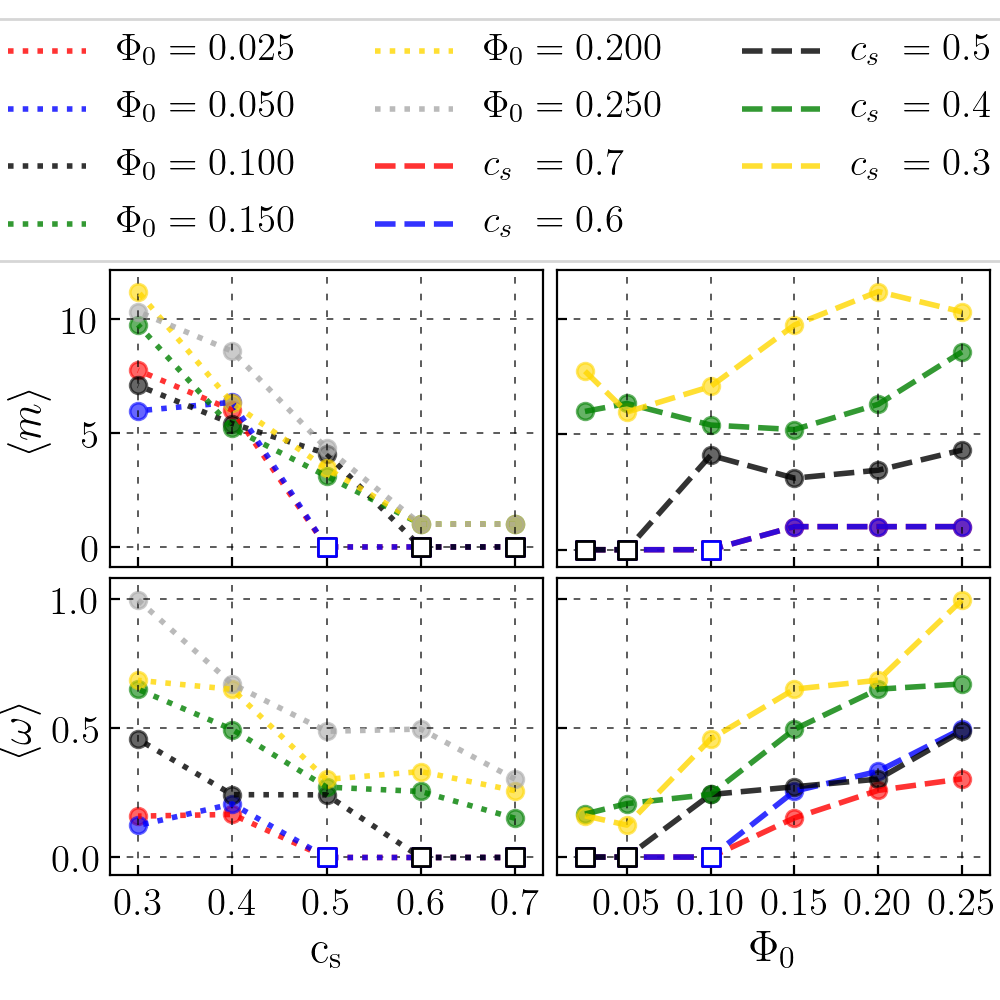}
\includegraphics[width=\columnwidth]{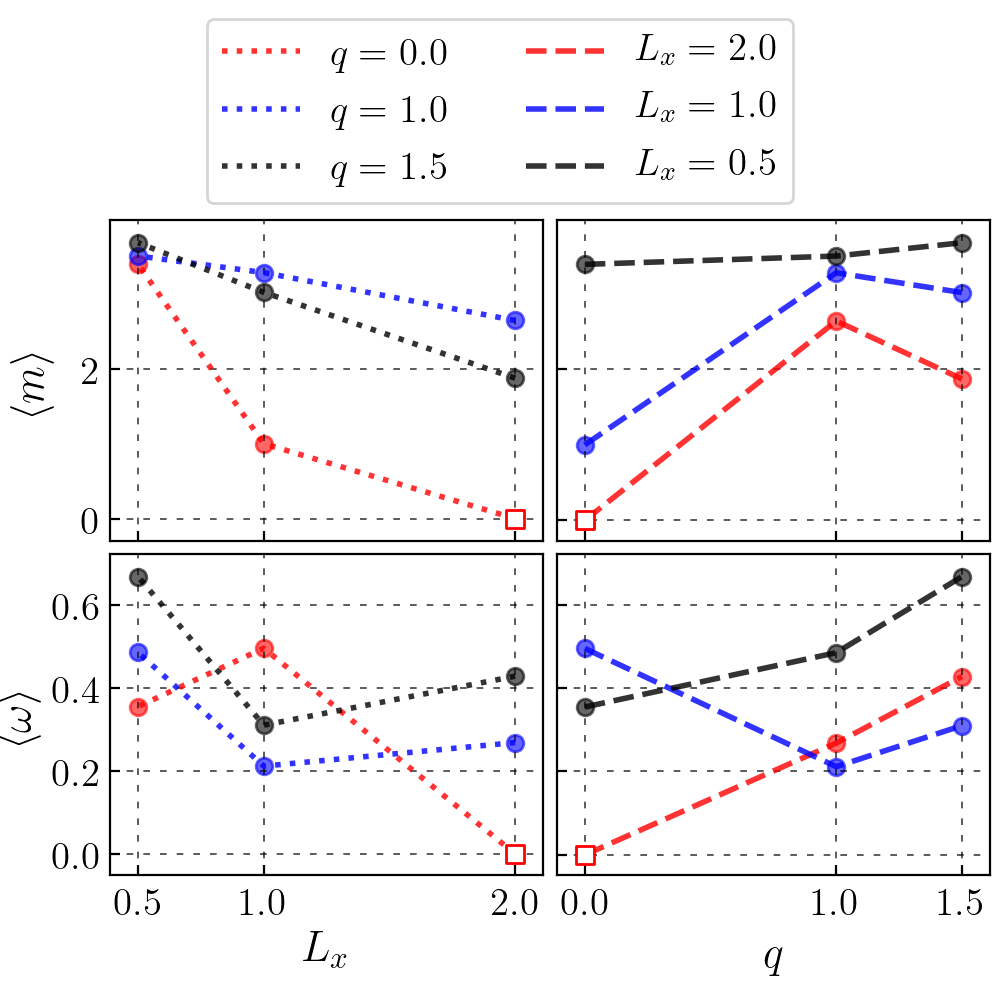}\vspace{-4mm}
\caption{The average growth rate $\langle\omega\rangle$ (Eq.~\ref{eq:meanomega}) and wavenumber $\langle m\rangle$ (Eq.~\ref{eq:meanm}) as a function of the various parameters for the first set of simulations (top) and the second set of simulations (bottom). Empty squares correspond to stable simulations, to which we assign $\langle \omega \rangle = \langle m \rangle = 0$.}
\label{fig:trends1}
\end{figure}

\begin{figure*}
\includegraphics[width=0.6\textwidth]{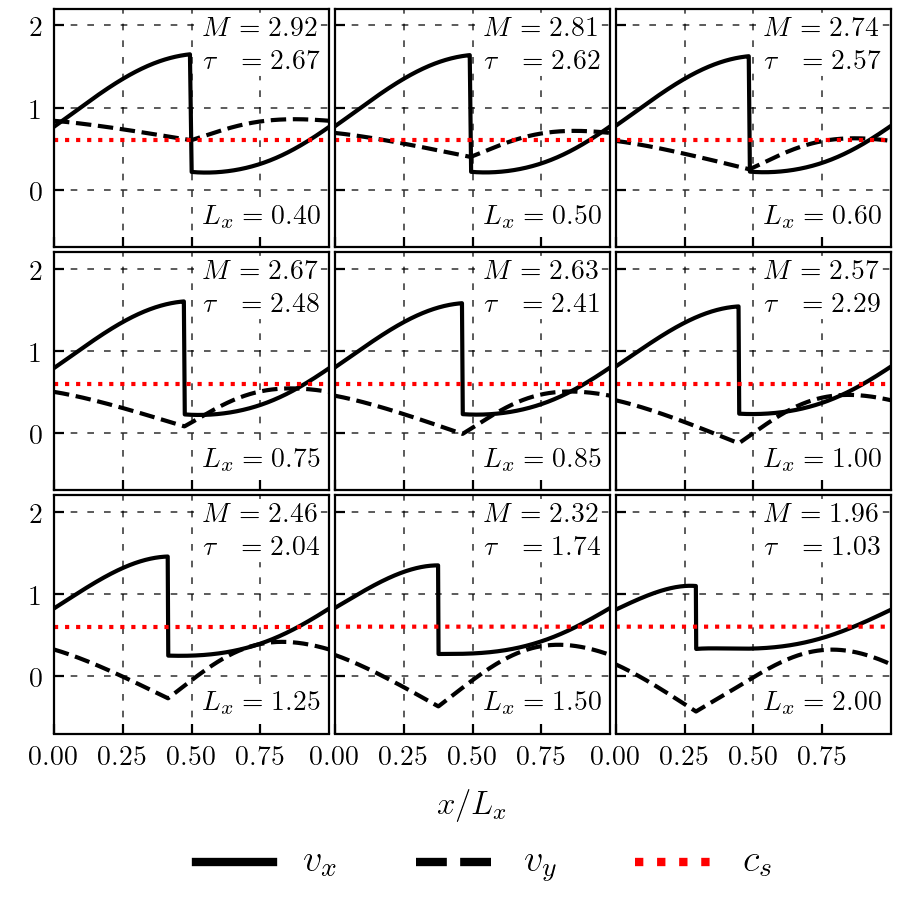}
\includegraphics[width=\textwidth]{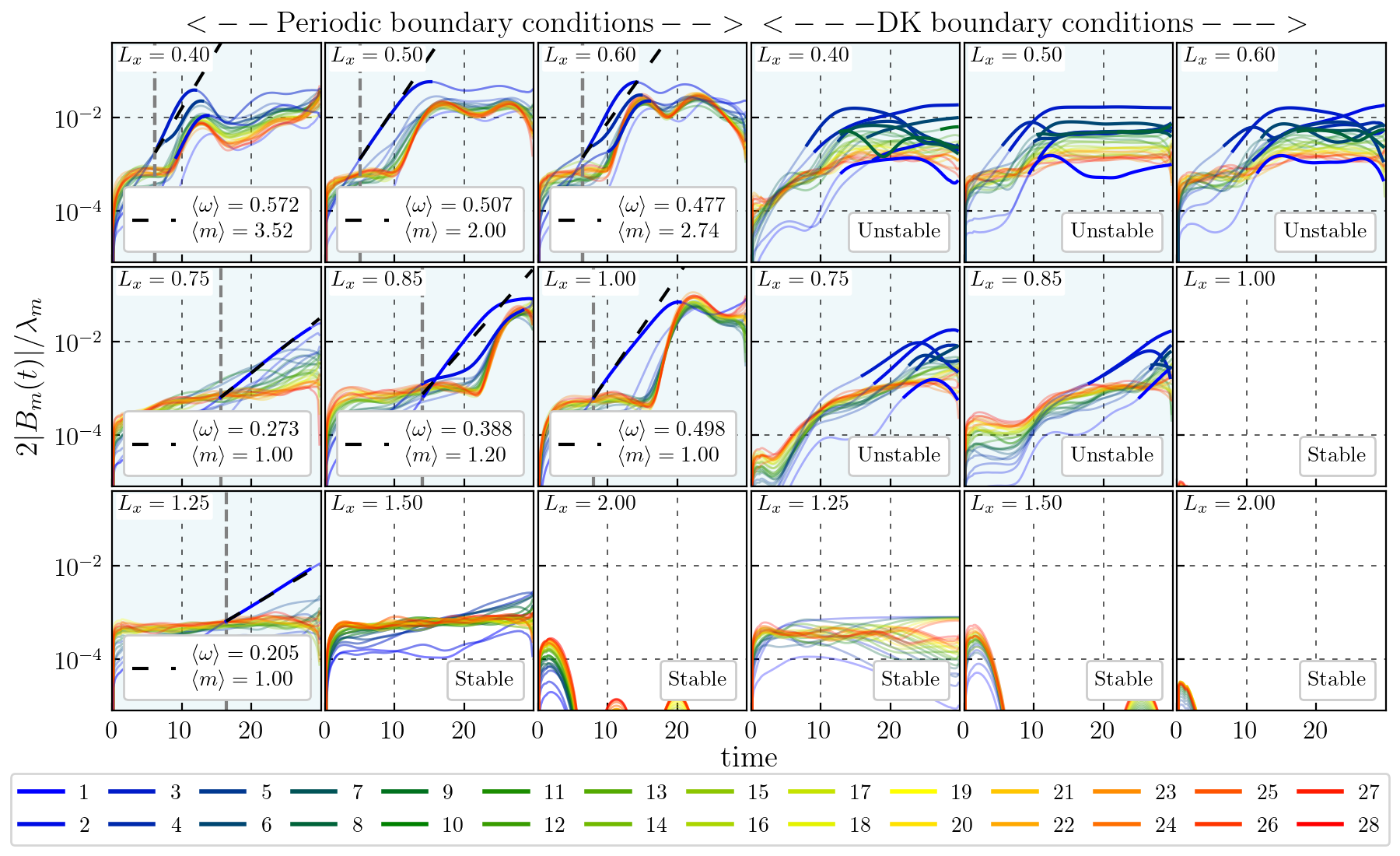}
\caption{Same as Fig.~\ref{fig:scanset2}, but for the third set of simulations.}
\label{fig:scanset3}
\end{figure*}

\begin{figure}
\centering
\includegraphics[width=\columnwidth]{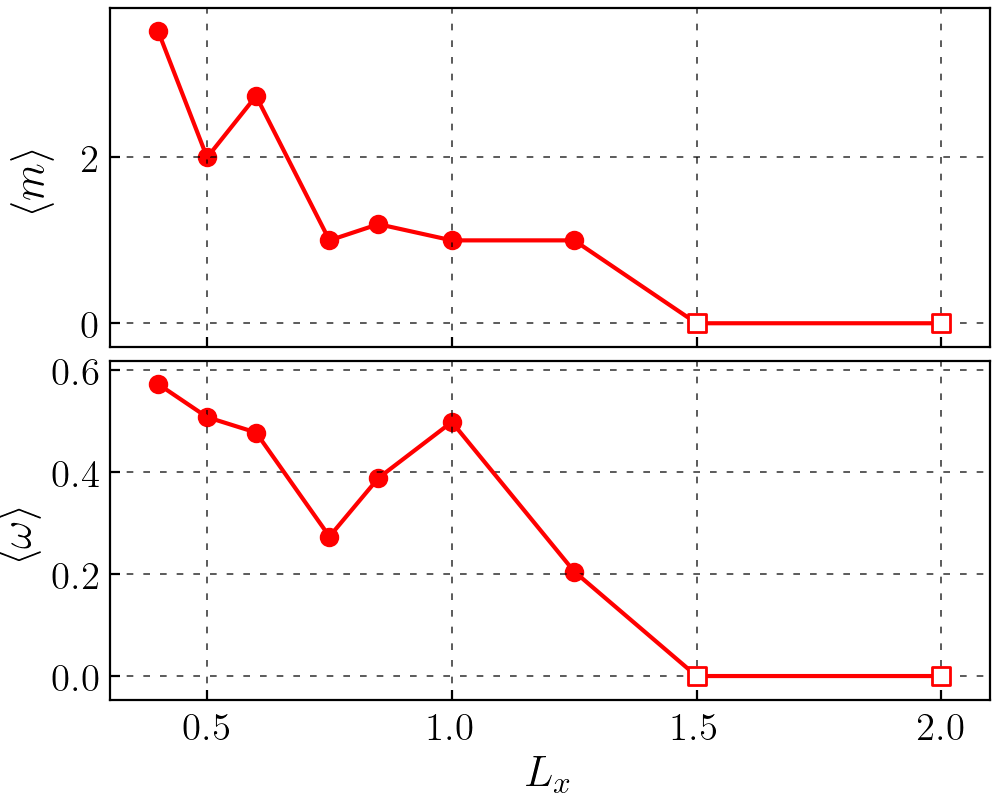}
\caption{The average growth rate $\langle\omega\rangle$ (Eq.~\ref{eq:meanomega}) and wavenumber $\langle m\rangle$ (Eq.~\ref{eq:meanm}) as a function of interarm distance for the third set of simulations. Empty squares correspond to stable simulations, to which we assign $\langle \omega \rangle = \langle m \rangle = 0$.}
\label{fig:trends2}
\end{figure}

\begin{figure*}
\centering
\includegraphics[width=\textwidth]{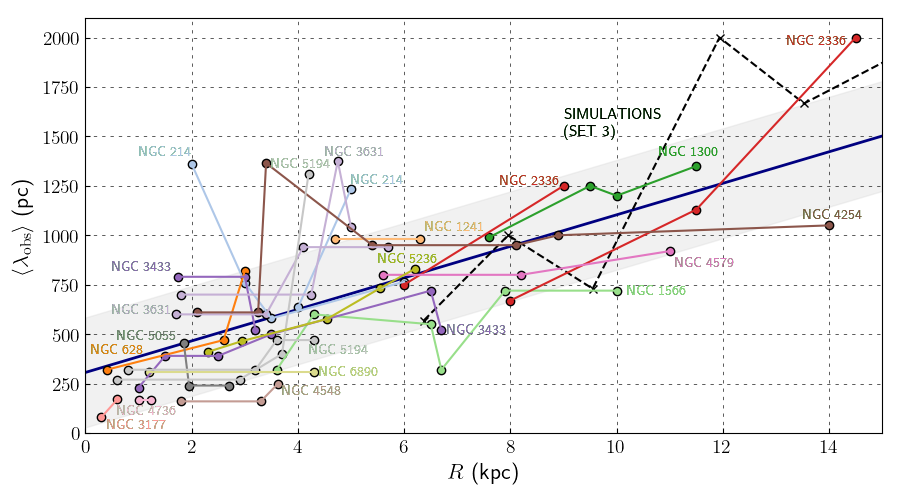}
\caption{Separation of feathers/spurs as a function of galactocentric radius $R$ for a sample of 20 galaxies in the survey of \citet{LaVigne2006} (see Table~\ref{tab:realobservations}). The grey band indicates a linear fit to the data, and the thickness is the standard deviation. The dashed line indicates $\langle \lambda \rangle$ for simulation set 3 (Table \ref{tab:scan}) assuming a pitch angle of $\sin(i)=0.02$ to convert from interarm separation $L_x$ to galactocentric radius $R$ (see Eq.~\ref{eq:L}), and using the scalings discussed in Section~\ref{sec:params} to convert from dimensionless to physical units. Changing the value of $\sin(i)$ changes the slope of the curve, while changing the scalings enlarges or shrinks the curve without affecting its slope.}
\label{fig:Trends_LV06}
\end{figure*}

The wiggle instability is an instability of the shock front. We, therefore, track the shock front in the simulations as a function of time, and we analyse its evolution by performing a Fourier decomposition of its shape.

Let us call $f(y,t)$ the $x$-displacement of the shock front with respect to its equilibrium position. We detect the shock front in each snapshot using the large density jump that characterises it. We estimate the density gradient $\pa_x \rho$ using finite differences along each horizontal slice, and we define the shock position to be the point where $|\pa_x \rho|$ is maximum. In this way, we obtain the position of the shock front $f(y=y_n,t_i)$ at each grid point $y=y_n$ and time $t=t_i$. This simple method tracks well the shock position as a function of time (see Figure~\ref{fig:examples}). 

We analyse the shape of the shock $f(y,t)$ for fixed $t$ using a discrete Fourier transform: 
\begin{equation} \label{eq:fourier}
f(y_n,t) = \sum_{m=0}^{N_y-1}B_m(t) \exp(i k_m y_n),
\end{equation}
where the wavenumber is
\begin{equation}
k_m = \frac{2\pi m}{L_y}\,, \quad m \in \{0,1,\dots,N_y-1\}
\end{equation}
and 
\begin{equation}
y_n = \frac{n L_y}{N_y}\,, \quad n \in \{0,1,\dots,N_y-1\}
\end{equation}
and $B_m=\bar{B}_{N_y-m}$, where the bar indicates the complex conjugate, because $f$ is real. Note that due to the finite size of the computational box in the $y$ direction, only discrete values of the wavenumber $k_m$ and of the wavelength $\lambda_m = 2\pi/k_m= L_y/m$, where $m$ is a positive integer, are allowed.

To analyse the temporal growth of the amplitudes, we smooth the $B_m(t)$ curves with a moving average of over 13 snapshots (corresponding to a total smoothing interval of $\Delta t =2.6$ in dimensionless units). This removes small-scale noise. Then we fit the smoothed curves with the following function: 
\begin{equation}  
B_{m,\mathrm{smooth}}(t) = a e^{- \omega_m t}, \label{eq:simplefit}
\end{equation}
where $a$ is a constant and $\omega_m$ is real. This approach neglects potential oscillations coming from the imaginary part of $\omega_m$, which cannot be detected reliably due to noise in our numerical setup and are washed out by our smoothing procedure. However, since these oscillations have by definition a zero net time-average, they do not affect our measurement of the long-term growth rates.

\section{Three example simulations} \label{sec:examples}

In this section, we analyse in detail three example simulations and compare them to predictions from linear stability analysis. The parameters of these simulations are chosen because they correspond to the cases for which a linear stability analysis is available \citep{Sormani2017}, and because, as it will be discussed below, they exemplify three main behaviours: (A) a system with a \emph{single dominant unstable mode} that becomes \emph{stable} with inflow-outflow boundary conditions; (B) a system with \emph{multiple unstable modes} that becomes \emph{stable} with inflow-outflow boundary conditions; (C) a system with \emph{multiple unstable modes} that remains \emph{unstable} with inflow-outflow boundary conditions. The parameters of these simulations are reported in Table \ref{tab:examples}.

\subsection{Comparison with Linear Stability Analysis} \label{sec:linear}
\begin{table}
\centering
\begin{tabular}{ccccc}
Simulation & \hspace{0.2cm} $\cs$ \hspace{0.2cm} & \hspace{0.2cm} $\Phi_0$ \hspace{0.2cm}  & \hspace{0.2cm} $L_x$ \hspace{0.2cm} & \hspace{0.2cm} $q$ \hspace{0.2cm} \\
\hline
A & 0.7 & 0.25  & 1.0 & 0 \\
B & 0.3 & 0.025 & 1.0 & 0 \\
C & 0.3 & 0.25  & 1.0 & 0 \\
\hline
\end{tabular}
\caption{Parameters of the example simulations analysed in Section~\ref{sec:examples}. Parameters are given in dimensionless form (see Section \ref{sec:params}).}
\label{tab:examples}
\end{table}

The left column in Figure \ref{fig:examples} shows the evolution of the surface density in the three simulations. At $t=0$ the shock front is a straight line in both simulations and then starts oscillating due to the wiggle instability. It is immediately evident that in simulation A (top) the instability is dominated by a single mode with a large wavelength ($m=1$), while in simulations B (middle) and C (bottom) there are multiple unstable modes.

The right column in Figure \ref{fig:examples} shows the evolution of individual modes as a function of time. The black dashed horizontal lines mark the values at which the amplitude becomes larger than the grid resolution, and the green shaded area indicates the region where instability is detectable, defined as the region where at least one mode is above the horizontal dashed line. Let us first consider in more detail simulation A. The first mode to cross the horizontal line is the $m=1$ at $t \simeq 12$. The subsequent evolution of this mode is very well approximated by exponential growth. At $t\simeq25$ the curve flattens and the growth stops as the instability saturates and we enter the non-linear regime. Other modes cross the line at $t\geq22$. The $m\geq 2$ modes also appear to grow exponentially after crossing the black dashed line. However, as we discuss below, we believe that the growth of the $m>1$ modes is driven by non-linear coupling between them and the $m=1$ mode.

The evolution of the amplitudes in simulation B is more complex. The first mode to cross the black dashed line is $m=8$ at $t\simeq18$, followed shortly by $m=4,7,11,10$. Of these, modes $m=10$ and $11$ saturate very quickly, while modes $8,4,7$ grow exponentially for some time before saturating. The morphology of the system shows the typical plumes of the wiggle instability, easily identifiable by visual inspection in the left column of Fig.~\ref{fig:examples}. Simulation C is very strongly unstable and shows the fastest evolution of the amplitudes. Multiple modes have already crossed the black dashed ``instability'' line at $t=0.2$.

Figure \ref{fig:dispersionrelation} compares the growth rates of the modes in the three simulations, measured by fitting Equation~\ref{eq:simplefit} to the amplitudes in Figure~\ref{fig:examples}, with rates predicted by the linear stability analysis of \citet{Sormani2017}. In the linear analysis, all values of $k_m$ are allowed, while in the simulations only discrete values $k_m=2\pi m/L_y$ where $m$ is an integer are allowed because of the finite size of the box in the $y$ direction. We measure the growth rate only for modes that start to grow exponentially immediately after the beginning of the green shaded area (which indicates when the amplitude of the first mode crosses the black dashed line). We argue below that the modes that start growing at later times are not genuinely unstable, because their growth is driven by the non-linear coupling between them and the genuinely unstable modes. We have verified that the measured growth rates do not depend significantly on the random seed of the initial noise.

The blue dot in the top panel of Figure~\ref{fig:dispersionrelation} shows the growth rate of the dominant $m=1$ for simulation A. The growth rate of this mode matches very well the one predicted by the linear stability analysis. The red squares indicate the growth rates for an additional simulation that is identical to simulation A except that we doubled the $y$ size of the simulation box, $L_y=4$, allowing for twice possible values of $k_m$. The growth rate of the first three modes of this simulation also matches very well those predicted by the linear analysis.

As mentioned above, the amplitude of the mode with $k_m = 2\pi$ in simulation A (see $m=2$ curve in the bottom-left of Figure~\ref{fig:examples}) also grows exponentially at $t>22$. This growth does not match the linear analysis, which predicts very slow growth (see $k_m=2\pi$ in the top panel of Fig.~\ref{fig:dispersionrelation}). To investigate the origin of this mismatch, we have run another simulation identical to simulation A with $L_y=1$. In this simulation, the smallest possible wavenumber is $k_m=2\pi$, while the $k_m = \pi$ mode is not possible because the box is not large enough. In this additional simulation, we do not observe any significant growth of the $k_m=2\pi$ mode. This suggests that the growth of the $k_m=2\pi$ mode in the simulation with $L_y=2$ is driven by its coupling to the $k_m=\pi$ mode. We have further confirmed this by running additional simulations with controlled initial conditions in which we excite only selected modes. We observe that if we excite only a certain wavenumber $k_m=k_0$, the growth of the modes whose wavenumber is an integer multiple of $k_0$ is enhanced. This confirms that there is significant non-linear interaction between the modes, even for relatively small values of the amplitudes ($|B_m|/\lambda_m\simeq1\%$). This coupling is neglected in the linear analysis, which assumes that each mode is independent.

The middle panel in Figure~\ref{fig:dispersionrelation} shows that simulation B, unlike simulation A, does not match very well the predictions of the linear analysis. The comparison is incomplete because the most unstable mode observed in simulation B corresponds to $k_m=8\pi$, which lies outside the range studied in the linear analysis of \citet{Sormani2017} ($k_m\leq 20$). Only one of the observed unstable modes ($k_m=4\pi$) is covered by the linear analysis in \citet{Sormani2017}. The growth rate of this mode in simulation B does not match very well the one predicted by the linear analysis. This is likely because the growth of the $m=4$ mode is affected by non-linear coupling to the $m=8$ mode since their wavelengths are multiple of each other. A less likely possibility is that an unstable mode has been missed in the linear analysis of \citet{Sormani2017}.

In the bottom panel of Figure~\ref{fig:dispersionrelation}, we see a drastic mismatch between our simulation C results and its corresponding predictions. All modes are unstable in this simulations, and contrary to its linear analysis predictions their growth rates are similar. We believe this is due to strong coupling between the unstable modes in the simulations, which makes them grow ``as a group''.

We conclude that the linear analysis works well only under very stringent conditions, i.e.\ when the evolution of the wiggle instability is dominated by a single unstable mode. Systems with multiple unstable modes show surprisingly strong non-linear coupling between modes, even for relatively low amplitudes ($|B_m|/\lambda_m\simeq1\%$). This coupling between modes invalidates the linear analysis, which is performed under the assumption that the modes are independent.

\subsection{Physical origin of the wiggle instability and the impact of boundary conditions} \label{sec:KH}

The boundary conditions are critical for the development of the wiggle instability, as has been already emphasised by \citet{Kim32014} and \citet{Sormani2017}. To understand why, let us briefly discuss the stability of shock fronts in general. The stability of shocks with respect to the formation of ``ripples'' and ``corrugations'' on their surface was studied in the classic work of \citet{Dy1954} and \citet{Kont1958} (see also \S90 in \citealt{Landau}). The unanimous conclusion of these works was that shock fronts are essentially always stable, except under exotic circumstances \citep{Landau}. However, these works made one key assumption: the pre-shock flow is unperturbed because the supersonic velocity of the pre-shock flow does not allow any signal to travel upstream. 

While the assumption of unperturbed pre-shock flow is appropriate for most applications, it is not appropriate for galactic spiral shocks, because the gas leaving one spiral arm will later pass through the next spiral arm(s). Thus, the gas upstream of the shock is not necessarily unperturbed: it can contain perturbations coming from the previous spiral arms. This seemingly innocuous difference can drastically change the conclusion about the stability of shocks. Indeed, an initial perturbation can be greatly amplified when passing through a shock if it resonates with the natural oscillation frequencies of the shock front \citep{Dy1954,Kont1958,McKenzie1968}. \citet{Kim32014} and \citet{Sormani2017} argued that the amplification of perturbations through successive shock passages is what gives rise to the wiggle instability. In particular, \citet{Sormani2017} used a linear stability analysis to show that some systems are stable with inflow-outflow boundary conditions (akin to the classic works above), but are unstable with periodic boundary conditions (which mimic the presence of multiple shocks in succession), proving that in these systems it is the amplification of perturbation in multiple shock passages that causes the instability. One of the goals of this paper is to test this, using simulations.

Figure~\ref{fig:periodicvsdk} shows what happens in the three simulations if we switch from periodic (left panels) to inflow-outflow (right panels) boundary conditions. Simulations A and B display wiggle instability with periodic boundary conditions but become completely stable with inflow-outflow boundary conditions (the shock front shows no signs of evolution). This proves the KHI is \emph{not} responsible for the wiggle instability in simulations A and B. Indeed, if the KHI were responsible for the wiggle instability in these systems, it would not disappear by changing the boundary conditions away from the post-shock region, where the shear is highest. The wiggle instability in simulations A and B is therefore caused exclusively by the amplification of perturbations through multiple shock passages (item ii in Sect.\ref{sec:history}). This also explains why early theoretical studies \citep{Nelson1977,Bal1985,Balbus1988,Dwarkadas1996} found spiral shocks to be stable: they used boundary conditions akin to the inflow-outflow boundary conditions used here, which are however not appropriate to study the stability of spiral shocks in global simulations such as those of \citet{Wada04}.

Simulation C shows different behaviour. This simulation is unstable with \emph{both} periodic and inflow-outflow boundary conditions. The instability develops \emph{before} the gas has had time to cross the simulation box in the $x$ direction. This proves that the wiggle instability cannot be due to successive shock passages in this case, since there have not been multiple shock passages. It seems likely that the wiggle instability in simulation C is instead caused by KHI from the very high shear $\tau=\pa_x v_y$ present immediately after the shock. Note that, consistent with this interpretation, the instability develops much faster than in Simulations A and B because the KHI-driven wiggles do not need to wait for the material to complete one period in the $x$ direction to grow. 

The linear analysis of \citet{Sormani2017} correctly predicts the overall stability/instability of all three simulations with both types of boundary conditions (i.e.\ whether the system as a whole is stable or unstable), but for simulations B and C it fails to quantitatively predict the growth rate of the unstable modes (Sect.~\ref{sec:linear}).

To summarise, we have proven that two distinct physical origins are possible for the wiggle instability, depending on the underlying parameters. The wiggle instability in simulations A and B is purely caused by the amplification of perturbation through successive shock passages. The wiggle instability in simulation C is primarily due to KHI caused by the large post-shock shear. We will see in Section~\ref{sec:scan} that the amount of post-shock shear in the steady-state solutions is indeed a very good predictor of whether the system is subject to KHI-driven wiggles.

\section{Parameter space scan} \label{sec:scan}
 
We explore the parameter space by running three sets of simulations. In the first set, we vary the sound speed and the spiral potential strength. In the second we vary the shear factor and the interarm separation. In the third, we look in more detail at the effects of varying the interarm separation. The parameters of all simulations are listed in Table \ref{tab:scan}. The three simulations analysed in Sect.~\ref{sec:examples} are part of the first simulation set and are highlighted in the table.

We quantify the properties of the wiggle instability using two main quantities. The first is the mean average wavenumber of the unstable modes, defined by:
\begin{equation} \label{eq:meanm}
\langle m\rangle = \frac{1}{N} \sum_{i=0}^{N} \langle m \rangle_i\,,
\end{equation}
where
\begin{equation}
   \langle m \rangle_i = \frac{\sum_{m=1}^{30} m B_m(t_i)}{\sum_{m=1}^{30} B_m(t_i)} \,,
\end{equation}
is the average wavenumber at time $t_i$, weighted by the amplitude of the various modes. The sum over $i$ is extended over $t_i = \{t_0,t_0 + i \Delta t,t_0 + 2 \Delta t, \dots, t_0 + N \Delta t\}$ where $\Delta t = 0.02$ in dimensionless units is the time interval between snapshots, and $t_0$ is the earliest time at which one mode (which we call $m_0$) becomes greater than the grid resolution (i.e.\ the beginning of the green shaded area in Figure \ref{fig:examples}). We take $N$ to be the nearest integer to $t_{\rm est}/\Delta t$, where $t_{\rm est}=1/\omega_{m_0}$ is an estimate of the typical growth time of the instability obtained using the instantaneous growth rate $\omega_{m_0}$ of the $m_0$ mode at $t=t_0$. Typical values of $t_{\rm est}$ are in the range $1$-$10$ in dimensionless units. 

A quantity closely related to $\langle m \rangle$ is the average wavelength of the unstable modes (see Section \ref{sec:shock}):
\begin{equation} \label{eq:meanlambda}
    \langle \lambda \rangle = \frac{L_y}{\langle m \rangle} \,.
\end{equation}
The value of $\langle \lambda \rangle$ is useful for comparison with observations because it characterises the average spacing between feathers/spurs generated by the wiggle instability.

The second quantity we define is the average growth rate:
\begin{equation} \label{eq:meanomega}
\langle \omega \rangle = \frac{1}{N} \sum_{i=0}^{N} \langle \omega \rangle_i \,,
\end{equation}
where
\begin{equation}
   \langle \omega \rangle_i = \frac{\sum_{m=1}^{30} \omega_m B_m(t_i)}{\sum_{m=1}^{30} B_m(t_i)}  \,,
\end{equation}
and $\omega_m$ is growth rate of mode $m$ obtained by fitting Equation \eqref{eq:simplefit} to the smoothed $B_m(t)$ curves in the range $[t_0,t_0+t_{\rm est}]$. A related quantity is $\langle T_{\rm grow} \rangle = 2\pi/\langle \omega \rangle$, which quantifies the timescale over which the instability grows. 

The values of $\langle m \rangle$ and $\langle \omega \rangle$ for each simulation under periodic boundary conditions are listed in Table \ref{tab:scan}.

\subsection{Sound speed and spiral potential strength}\label{psscp}

Figure~\ref{fig:ic1} shows the initial condition for the first set of simulations, in which we vary the sound speed $\cs$ and the spiral potential strength $\Phi_0$ while keeping fixed the other parameters (see Table \ref{tab:scan}). Figures \ref{fig:cpp} and \ref{fig:cpdk} show the surface density at a later time for periodic and inflow-outflow (DK) boundary conditions respectively. Most of the simulations with periodic boundary conditions display the wiggle instability (Figs.~\ref{fig:cpp} and \ref{fig:ftcpp}), and those that do not exhibit the wiggle instability would probably develop it if the simulation were continued for longer times. Comparing Figs.~\ref{fig:cpp} and \ref{fig:ftcpp} with Figs.~\ref{fig:cpdk} and \ref{fig:ftcpdk} shows that some systems become stable when switching to inflow-outflow boundary conditions, while others (especially those with high $\Phi_0$/low $\cs$) remain unstable for inflow-outflow boundary conditions. In the former, the wiggle instability originates purely from the amplification of perturbation at successive shocks, while in the latter it is driven, at least in part, by the KHI (see Section~\ref{sec:KH}). Table \ref{tab:scan} shows that the post-shock shear $\tau$ is a good predictor of whether the simulation will display KHI-driven wiggles or periodicity-driven wiggles.

Figures \ref{fig:ftcpp} and \ref{fig:ftcpdk} show the time evolution of the Fourier modes for the case of periodic and inflow-outflow (DK) boundary conditions respectively. The top panels in Figure~\ref{fig:trends1} show the average wavenumber and growth rates of the unstable modes for the simulations with periodic boundary conditions. We can see the following trends:
\begin{enumerate}[leftmargin=*]
\item Simulations with higher $\Phi_0$ and lower $\cs$ tend to be more unstable. 
\item The average wavenumber $\langle m \rangle$ of the unstable modes increases for decreasing $\cs$ (see Fig.~\ref{fig:trends1}), while showing an extremely weak dependence on $\Phi_0$.
\item The average growth rate $\langle \omega \rangle$ increases for increasing $\Phi_0$ and decreasing $\cs$ (see Fig.~\ref{fig:trends1}).
\end{enumerate}
Finally, note that modes within the same simulation tend to saturate to similar values of the normalised amplitudes $B_m/\lambda_m$ (Fig.~\ref{fig:ftcpp}).

\subsection{Shear factor} \label{sec:PSS_}
In the second set of simulations, we analyse the dependence of the wiggle instability on the term containing $q$ in Equation~\ref{eq:2e1}, which arises due to differential rotation in the galaxy. The value $q=0$ corresponds to solid-body rotation, $q=1$ to a flat rotation curve, and $q=1.5$ to Keplerian rotation. Figure~\ref{fig:scanset2} shows the amplitude of the Fourier modes as a function of time for both periodic and inflow-outflow boundary conditions, while the bottom part of Figure~\ref{fig:trends1} shows the average wavenumber and growth rate as a function of $q$. From these figures, we conclude that the wiggle instability depends only weakly on the shear factor $q$, although decreasing $q$ tends to stabilise the system.

\subsection{Spiral arms separation}\label{sec:PSS_L}

In the third set of simulations, we examine the instability as a function of the interarm separation $L_x$. Figure~\ref{fig:scanset3} shows the evolution of the Fourier amplitudes for the second and third set of simulations as a function of time, and Figure~\ref{fig:trends2} shows the average wavenumber and growth rate of the unstable modes. We note that:
\begin{enumerate}[leftmargin=*]
\item The growth rate $\langle \omega \rangle$ tends to decrease for increasing $L_x$. This is expected in the case that the wiggle instability originates from the amplification of perturbation at multiple shocks since for larger $L_x$ the time interval between consecutive shock passages is larger (disturbances travel longer distances in the $x$ direction to reach the next shock).
\item The average wavenumber $\langle m \rangle$ decreases as we increase $L_x$. We attribute this effect to the longer travel times between shocks which promote the dilution/dispersion of small distortions corresponding to large wavenumbers.
\item For $L_x\geq1.5$ the growth is too slow for the wiggle instability to have any impact on real galaxies ($\langle T_{\rm grow}\rangle \gtrsim 1.5 \Gyr$ in physical units).
\end{enumerate}
Finally, note that simulations with large $L_x$ may appear stable because of the finite length $L_y$, which does not allow large wavelengths that may dominate the instability in these cases (see also discussion in Section~\ref{sec:examples}).

\begin{table*}
\begin{center}
\caption{Simulations in the parameter space scan. (1) ID: identification number. (2) $\cs$: dimensionless sound speed. (3) $\Phi_0$: dimensionless spiral potential strength. (4) $L_x$: dimensionless interarm separation. (5) $q$: shear factor. (6) $M=v_x/\cs$: Mach number of the steady-state solution,  where $v_x$ is the velocity on the upstream side of the shock (i.e., immediately before the shock). (7) $T_x=\int_0^{L_x} \, \di x/v_x$: horizontal period of the steady-state solution, i.e.\ time required for the $x$ coordinate of a gas parcel to return to its initial value. (8) $\tau=\pa_x v_y$: post-shock shear in the steady-state solution, calculated using the velocity on the downstream side of the shock (i.e., immediately after the shock). (9) $\langle \omega \rangle$: average growth rate of the unstable modes (Eq.~\ref{eq:meanomega}). (10) $\langle m \rangle$: average unstable wavenumber (Eq.~\ref{eq:meanm}). (11) $\langle\lambda\rangle = L_y/\langle m\rangle$: average unstable wavelength (Eq.~\ref{eq:meanlambda}) in pc, converted to physical units using the typical physical scalings from Section~\ref{sec:params}. (12) KHI: presence of the Kelvin-Helmholtz instability, i.e. whether the simulation is unstable under inflow-outflow boundary conditions. $\langle\omega\rangle$, $\langle m \rangle$ and $\langle \lambda \rangle$ are calculated for periodic boundary conditions.}
\label{tab:scan}
\begin{tabular}{l|cccc|ccc|cccc}
ID  & $\cs$ &  $\Phi_0$   &  $L_x$ &  $q$ & $M$ & $T_x$ & $\tau$ & $\langle\omega\rangle$ & $\langle m\rangle$ & $\langle \lambda \rangle$ (pc) & KHI \\
\hline
\multicolumn{12}{c}{First set of simulations}  \\
\hline
01 (simulation B) & 0.3 & 0.025& 1.0 & 0 & 2.01  & 2.001 & 3.98   & 0.160 & 7.74  & 258 & No  \\
02 & 0.4 & 0.025& 1.0 & 0 & 1.68  & 1.999 & 1.98   & 0.167 & 5.99  & 334 & No  \\
03 & 0.5 & 0.025& 1.0 & 0 & 1.36  & 1.999 & 0.67   & ---   & ---   & --- & No  \\
04 & 0.6 & 0.025& 1.0 & 0 & ---   & ---   & ---    & ---   & ---   & --- & --- \\
05 & 0.7 & 0.025& 1.0 & 0 & ---   & ---   & ---    & ---   & ---   & --- & --- \\
06 & 0.3 & 0.05 & 1.0 & 0 & 2.550 & 1.999 & 5.555  & 0.124 & 5.96  & 335 & Yes \\
07 & 0.4 & 0.05 & 1.0 & 0 & 2.640 & 1.997 & 2.848  & 0.208 & 6.34  & 316 & No  \\
08 & 0.5 & 0.05 & 1.0 & 0 & 1.668 & 2.000 & 1.247  & ---   & ---   & --- & No  \\
09 & 0.6 & 0.05 & 1.0 & 0 & 1.318 & 2.000 & 0.190  & ---   & ---   & --- & No  \\
10 & 0.7 & 0.05 & 1.0 & 0 & ---   & ---   & ---    & ---   & ---   & --- & --- \\
11 & 0.3 & 0.10 & 1.0 & 0 & 3.322 & 1.998 & 7.911  & 0.458 & 7.07  & 283 & Yes \\
12 & 0.4 & 0.10 & 1.0 & 0 & 2.635 & 1.996 & 4.217  & 0.243 & 5.40  & 370 & Yes \\
13 & 0.5 & 0.10 & 1.0 & 0 & 2.133 & 1.999 & 2.152  & 0.243 & 4.09  & 489 & No  \\
14 & 0.6 & 0.10 & 1.0 & 0 & 1.731 & 1.999 & 0.861  & ---   & ---   & --- & No  \\
15 & 0.7 & 0.10 & 1.0 & 0 & 1.33  & 2.001 & -0.08  & ---   & ---   & --- & No  \\
16 & 0.3 & 0.15 & 1.0 & 0 & 3.945 & 2.002 & 9.909  & 0.652 & 9.73  & 206 & Yes \\
17 & 0.4 & 0.15 & 1.0 & 0 & 3.097 & 1.998 & 5.373  & 0.496 & 5.20  & 384 & Yes \\
18 & 0.5 & 0.15 & 1.0 & 0 & 2.504 & 2.000 & 2.904  & 0.273 & 3.10  & 645 & No  \\
19 & 0.6 & 0.15 & 1.0 & 0 & 2.050 & 2.000 & 1.398  & 0.256 & 1.00  & 2000& No  \\
20 & 0.7 & 0.15 & 1.0 & 0 & 1.65  & 2.002 & 0.37   & 0.152 & 1.00  & 2000& No  \\
21 & 0.3 & 0.20 & 1.0 & 0 & 4.495 & 2.003 & 11.617 & 0.687 & 11.18 & 179 & Yes \\
22 & 0.4 & 0.20 & 1.0 & 0 & 3.504 & 1.995 & 6.353  & 0.652 & 6.30  & 318 & Yes \\
23 & 0.5 & 0.20 & 1.0 & 0 & 2.828 & 1.999 & 3.563  & 0.303 & 3.45  & 580 & Yes \\
24 & 0.6 & 0.20 & 1.0 & 0 & 2.324 & 1.999 & 1.868  & 0.333 & 1.00  & 2000& No  \\
25 & 0.7 & 0.20 & 1.0 & 0 & 1.91  & 2.000 & 0.75   & 0.259 & 1.00  & 2000& No  \\
26 (simulation C) & 0.3 & 0.25 & 1.0 & 0 & 5.001 & 1.999 & 12.919 & 0.998 & 10.30 & 194 & Yes \\
27 & 0.4 & 0.25 & 1.0 & 0 & 3.880 & 1.999 & 7.298  & 0.672 & 8.57  & 233 & Yes \\
28 & 0.5 & 0.25 & 1.0 & 0 & 3.124 & 1.995 & 4.154  & 0.489 & 4.32  & 463 & Yes \\
29 & 0.6 & 0.25 & 1.0 & 0 & 2.571 & 1.996 & 2.295  & 0.498 & 1.00  & 2000& No  \\
30 (simulation A) & 0.7 & 0.25 & 1.0 & 0 & 2.13  & 1.998 & 1.07   & 0.304 & 1.00  & 2000& No  \\
 \hline
 \multicolumn{12}{c}{Second set of simulations} \\
 \hline
01 & 0.6 & 0.25 & 0.5 & 0   & 2.809 & 0.998 & 2.621 & 0.355 & 3.39 & 590 & Yes \\
02 & 0.6 & 0.25 & 1.0 & 0   & 2.571 & 1.996 & 2.295 & 0.495 & 1.00 & 2000& No  \\
03 & 0.6 & 0.25 & 2.0 & 0   & 1.960 & 3.999 & 1.033 & ---   & ---  & --- & No  \\
04 & 0.6 & 0.25 & 0.5 & 1.0 & 2.882 & 0.999 & 1.332 & 0.485 & 3.50 & 571 & No  \\
05 & 0.6 & 0.25 & 1.0 & 1.0 & 2.682 & 1.999 & 1.248 & 0.211 & 3.28 & 610 & No  \\
06 & 0.6 & 0.25 & 2.0 & 1.0 & 2.342 & 3.998 & 0.924 & 0.268 & 2.64 & 758 & No  \\
07 & 0.6 & 0.25 & 0.5 & 1.5 & 2.923 & 0.997 & 0.673 & 0.667 & 3.68 & 543 & No  \\
08 & 0.6 & 0.25 & 1.0 & 1.5 & 2.750 & 1.998 & 0.648 & 0.310 & 3.02 & 662 & No  \\
09 & 0.6 & 0.25 & 2.0 & 1.5 & 2.566 & 3.999 & 0.571 & 0.428 & 1.87 & 1070& No  \\
 \hline
 \multicolumn{12}{c}{Third set of simulations} \\
 \hline
01 & 0.6 & 0.25 & 0.40 & 0 & 2.924 & 0.799 & 2.665 & 0.572 & 3.52 & 568 & Yes  \\
02 & 0.6 & 0.25 & 0.50 & 0 & 2.809 & 0.998 & 2.621 & 0.507 & 2.00 & 1000& Yes  \\
03 & 0.6 & 0.25 & 0.60 & 0 & 2.739 & 1.199 & 2.567 & 0.477 & 2.74 & 730 & Yes  \\
04 & 0.6 & 0.25 & 0.75 & 0 & 2.667 & 1.499 & 2.481 & 0.273 & 1.00 & 2000& Yes  \\
05 & 0.6 & 0.25 & 0.85 & 0 & 2.628 & 1.700 & 2.411 & 0.388 & 1.20 & 1667& Yes  \\
06 & 0.6 & 0.25 & 1.00 & 0 & 2.571 & 1.996 & 2.295 & 0.498 & 1.00 & 2000& No   \\
07 & 0.6 & 0.25 & 1.25 & 0 & 2.462 & 2.499 & 2.044 & 0.205 & 1.00 & 2000& No   \\
08 & 0.6 & 0.25 & 1.50 & 0 & 2.324 & 3.000 & 1.738 & ---   & ---  & --- & No   \\
09 & 0.6 & 0.25 & 2.00 & 0 & 1.960 & 3.999 & 1.033 & ---   & ---  & --- & No   \\
\hline
\end{tabular}
\end{center}
\end{table*}

\section{Discussion} \label{sec:discussion}

\subsection{Comparison with observations}

We can compare our simulations to observations of spurs/feathers in disc galaxies. Figure \ref{fig:Trends_LV06} shows the spacing between feathers along the arm for a sample of 20 galaxies in the survey of \citet{LaVigne2006}. \citet{LaVigne2006} provide values for only two galaxies (NGC 3433 and
NGC 5985). To increase the number of data points, we manually measured the feather separation for 20 galaxies directly from the figures in \citet{LaVigne2006} by superimposing a Cartesian grid onto them. We corrected for the effect of inclination, although this has a minor impact on the results since galaxies with low inclinations are selected (see Table~2 in \citealt{LaVigne2006}). We have checked that for NGC 3433 and NGC 5985 our values are consistent with those provided by \citet{LaVigne2006}  within the errors. Our values are also listed in Table~\ref{tab:realobservations}.

Comparing values of $\langle \lambda \rangle$ from Table \ref{tab:scan} to those of $\langle \lambda_{\rm obs} \rangle$ in Table \ref{tab:realobservations}, we can see that the wiggle instability is able to reproduce the range of spacing seen in observations ($200$-$2000$ pc). However, it is not straightforward to make a correspondence between parameters of the simulations and observations because (i) there are multiple degeneracies between parameters, so that for a given $\langle \lambda_{\rm obs} \rangle$ there is no unique set of parameters that can reproduce it; (ii) as discussed in Section~\ref{sec:scan}, the predicted wavelength of the wiggle instability is extremely sensitive to parameters such as the sound speed and the spiral potential strength.

\cite{LaVigne2006} note that galaxies with weak spiral potential (class Sc and Sd) show decreased feather formation. This is consistent with our results where decreasing $\Phi_0$ decreases the strength of the instability.

Figure \ref{fig:Trends_LV06} shows that there is a clear correlation between feather spacing $\langle \lambda_{\rm obs} \rangle$ and galactocentric radius $R$ (the spacing increases with radius). Our simulations predict a similar trend, because $\langle \lambda\rangle$ increases with interarm distance $L_x$ (Figure~\ref{fig:trends2}), and the latter typically increases with $R$. To compare these two trends in more detail, we need to convert the interarm distance $L_x$ into galactocentric radius $R$. For the idealised logarithmic spiral arms used in this paper the two are related by (see Eq.~\ref{eq:L}):
\begin{equation} 
L_x = \left(\frac{2 \pi\sin(i)}{m_{\rm s}}\right) R\,,
\end{equation}
where $\sin(i)$ is the pitch angle and $m_{\rm s}$ is the number of spiral arms (not to be confused with the wavenumber). In this comparison we have the freedom to choose $\sin(i)$ arbitrarily, because specifying all the six parameters that characterise our idealised problem only fixes the product $v_{\rm c0} \sin(i)$ (see Section \ref{sec:params}). The value of $\sin(i)$ controls the predicted slope of the trend in the $\lambda-R$ plane. If we assume $m_{\rm s}=2$ and a typical pitch angle of $\sin(i)\simeq0.1$ \citep[e.g.][]{Savchenko2013} then our simulations predict a relation which is too steep (i.e., $\langle \lambda \rangle$ increases too quickly with $R$). The simulations can reproduce the observations well only if we assume an unrealistically small value of the pitch angle $\sin(i)=0.02$ (black dashed line in Figure \ref{fig:trends2}). 

However, the picture is complicated by several factors. First, the extreme sensitivity to the underlying parameters discussed above. It may be possible that the observed trend is better reproduced with a lower value of the sound speed than used in the third set of simulations. Second, the very idealised nature of our simulations. We assume that the gas is isothermal with constant sound speed (to which $\langle \lambda \rangle$ is very sensitive, as discussed in Section~\ref{sec:scan}). However, this is a very crude approximation because the interstellar medium is in reality a complex multi-phase medium, and it is not clear what effect this would have on the properties of the wiggle instability. We neglect self-gravity, magnetic fields, and stellar feedback, all of which are likely to affect $\langle \lambda \rangle$. We assume that spiral arms are long-lived and rigidly rotating with a definite pattern speed, while spiral arms in real galaxies are likely to be transient and time-dependent \citep[e.g.][]{Fujii2011,Wada2011,Sellwood2019}. In our local approximation, we also assume that the interarm separation remains constant and that conditions are strictly periodic, but in reality, there are variations between subsequent arm passages. Finally, we assume that the spirals have a small pitch angle ($\sin(i)\ll 1$), which is often not the case in real galaxies. Global simulations are probably essential for a full comparison of the wiggle instability to observations \citep{Wada04,Wada2008,Kim2014}.

We conclude that while the wiggle instability appears to be able to reproduce the typical range of feather spacing observed in real galaxies, the extreme sensitivity of the predicted spacing to the underlying parameters and the highly idealised nature of our setup makes a detailed comparison and constraining of the underlying parameters a difficult task.

\begin{table}
\caption{Properties of spurs/feathers for 20 galaxies in the Hubble Space Telescope Survey of \citet{LaVigne2006}. $\langle \lambda_{\rm obs} \rangle$: separation of feathers/spurs along the arm. $R$: galactocentric radius. }
\label{tab:realobservations}
\centering
\begin{tabular}{ccr}
\hline
Galaxy &  Type & $\langle \lambda_{\rm obs} \rangle$ in pc ($R$ in kpc) \\
\hline
NGC 0214 & SABRbc   & 635(4), 1235(5); 1360(2),         \\
&                   & 760(3), 580(3.5), 765(6)          \\
NGC 0628 & SASc     & 320(0.4), 470(2.6), 820(3)        \\
NGC 1241 & SBTb     & 980(4.7 - 6.3)                    \\
NGC 1300 & SBTbc    & 990(7.6), 1250(9.5),              \\
&                   & 1200(10), 1350(11.5)              \\
NGC 1566 & SABSbc   & 320(3.6), 600(4.3), 550(6.5),     \\
&                   & 320(6.7), 720(7.9 - 10)           \\
NGC 2336 & SABRbc   & 750(6), 1250(9); 670(8),          \\
&                   & 1130(11.5), 2000(14.5)            \\
NGC 3177 & SATb     & 80(0.3), 170(0.6)                 \\
NGC 3433 & SASC     & 230(1), 390(1.5 - 2.5),           \\
&                   & 500(3.5), 720(6.5), 520(6.7);     \\
&                   & 790(1.75 - 3), 520(3.2)           \\
NGC 3631 & SASc     & 600(1.7 - 3.4), 940(4.1 - 5.7),   \\
&               & 700(1.8 - 4.25), 1375(4.75), 1040(5)  \\
NGC 4254 & SASc     & 610(2.1 - 3.52), 1365(3.4),       \\
&               & 950(5.4 - 8.1), 1000(8.9), 1050(14)   \\
NGC 4548 & SBTb     & 160(1.8 - 3.3), 250(3.63)         \\
NGC 4579 & SABTb    & 800(5.6 - 8.2), 920(11)           \\
NGC 4736 & RSARab   & 167(1 - 1.3)                      \\
NGC 5055 & SATbc    & 456(1.85), 240(1.95 - 2.7)        \\
NGC 5194 & SASbcP   & 320(0.8 - 3.2), 470(3.6 - 4.3)    \\
&               & 270(0.6 - 2.9), 400(3.7), 1310(4.2)   \\
NGC 5236 & SABSc    & 410(2.3), 465(2.95), 577(4.55),   \\
&                   & 733(5.55), 830(6.2)               \\
NGC 6890 & SATb     & 310(1.2 - 4.3)                    \\
\hline
\end{tabular}
\end{table}

\subsection{Wiggle instability vs other mechanisms for the formation of spurs and feathers}

Numerous mechanisms have been proposed for spurs/feather formation. \cite{Kim2002,Kim2006} (see also \citealt{Lee2012,Lee2014}) proposed that spurs might form via a magneto-Jeans instability (MJI), in which magnetic fields favour the gravitational fragmentation and collapse of spiral arm crests by removing angular momentum from contracting regions. \cite{Kim2020} proposed that spurs/feathers originate from the stochastic accumulation of gas due to correlated supernova feedback. \cite{Dobb2006} proposed that spurs/feathers originate from the amplification of pre-existing perturbations at the arm crest, in a way that is reminiscent of one of the two possible physical origins of the wiggle instability.

One of the key properties of the wiggle instability is that it does not rely on the presence of magnetic fields, self-gravity and supernova feedback. Thus, feathering due to the wiggle instability should be present even in regions without much stellar feedback (e.g. the outer HI spiral arms in disc galaxies) or where the gas self-gravity is negligible. It might also be relevant in other contexts, such as spiral arms in protoplanetary discs \citep[e.g.][]{Rosotti2020}.

There has been some confusion regarding the relationship between the wiggle instability, the Kelvin-Helmholtz instability (KHI) and the instability that arises due to amplification of perturbations at multiple shock passages. In their original analysis, \cite{Wada04} proposed KHI as the origin of the wiggle instability, while \cite{Kim2014} and \cite{Sormani2017} attributed the wiggle instability to repeated passage of the gas through the shock-front. In this paper, we define wiggle instability as the phenomenon by which galactic spiral shocks form wiggles, ripples and corrugations on their surfaces. Using this definition, in Section~\ref{sec:KH} we have shown that the wiggle instability can originate from both the KHI and the amplification of perturbations at repeated shocks. These two distinct physical mechanisms can act alone or simultaneously, depending on the underlying parameters of the system.

Note that magnetic fields and the presence of supernova feedback all tend to suppress the wiggle instability \citep{Kim2006,Kim2015,Kim2020}. It's, however, challenging to determine the exact contribution of each mechanism. Observations of regions where one can exclude some of these mechanisms (e.g. where self-gravity or supernova feedback are negligible) might help in discriminating between them.

\section{Conclusion} \label{sec:conclusion}

We simulated a small patch of a typical spiral galaxy to study the wiggle instability in the simplest possible setup. We found the following results:
\begin{enumerate}[leftmargin=*]
\item We compared in detail the results of simulations with predictions from linear stability analysis by doing a Fourier decomposition of the perturbed shock front. The linear analysis works well only under specific circumstances, i.e.\ when the wiggle instability is caused by a single dominant unstable mode. When multiple unstable modes are present, they strongly couple and influence each other evolution even when the amplitudes are relatively small (ratio between the displacement of the shock front and the wavelength of the unstable mode $\sim 1 \%$. This non-linear coupling is not captured by the linear analysis. (Section~\ref{sec:examples})
\item The wiggle instability is physical and can have two distinct possible origins: the Kelvin-Helmholtz instability (item i in Sect.~\ref{sec:history}) or the amplification of perturbations at repeated shock passages (item ii in Sect.~\ref{sec:history}). The dominant mechanism depends on the underlying parameters. The KHI tends to be more important in systems with small sound speed $\cs$ and/or large spiral potential strength $\Phi_0$. (Section~\ref{sec:KH})
\item The properties of the wiggle instability are very sensitive to the underlying parameters, in particular the gas sound speed $\cs$, the spiral potential strength $\Phi_0$ and the interarm spacing $L_x$. The average separation of wiggle-driven spurs decreases with decreasing $\cs$ and/or decreasing $L_x$. The growth rate of the instability increases with increasing $\Phi_0$ and increases with decreasing $\cs$ and/or $L_x$ (Section~\ref{sec:scan}).
\item The wiggle instability can reproduce the range of spacing between feathers observed in real galaxies. However, the extreme sensitivity of the predicted spacing to the underlying parameters and the idealised nature of our setup makes constraining the underlying parameters a difficult task. Moreover, it is very challenging to disentangle the contribution of the wiggle instability from other mechanisms for substructure formation such as magneto-Jeans instabilities or correlated supernova feedback (Section~\ref{sec:discussion}).
\end{enumerate}

\section*{Acknowledgement}
%Y.M., M.C.S., E.S. and R.S.K.
We are grateful to the referee, Keiichi Wada, for a helpful and constructive report that improved the clarity of the paper. Y.M., M.C.S.\ and R.S.K.\ acknowledge support from the German Research Foundation (DFG) via the collaborative research centre (SFB 881, Project-ID 138713538) the Milky Way System (subprojects A1, B1, B2, and B8), from the Heidelberg Cluster of Excellence “STRUCTURES” in the framework of Germany’s Excellence Strategy (grant EXC2181/1, Project-ID 390900948), and from the European Research Council via the ERC Synergy Grant “ECOGAL” (grant 855130). The project benefited from computing resources provided by the State of Baden-W\"urttemberg through bwHPC and DFG through grant INST 35/1134-1 FUGG, and from the data storage facility SDS@hd supported through grant INST 35/1314-1 FUGG. We also acknowledge the Leibniz Computing Centre (LRZ) for providing HPC resources in project pr74nu.

\section*{Data availability}
The data underlying this article will be shared on reasonable request to the corresponding author. Videos of the simulations are available at \url{https://www.youtube.com/playlist?list=PLlsb6ZGKWbI77_XcdS8N87fJDXGei6Adh}.

\bibliography{bibliography}
\bibliographystyle{mnras}
\appendix

\section{Derivation of the basic equations}\label{sec:derivation}

Here we provide a detailed derivation of Equations \eqref{eq:2e1} and \eqref{eq:2e2}. We start by writing down the equations of fluid dynamics in a rotating frame. Then we introduce Roberts' spiral coordinate system, and rewrite the equations in this coordinate system without any approximation. Finally, we approximate the equations using the assumptions of locality and of small pitch angle.

\subsection{Fluid equations in a rotating frame} \label{sec:rotating}

The Euler and continuity equations in a frame rotating with pattern speed $\bfOmegap =\Omegap \hatez$ are:

\begin{align} 
& \pa_t \rho + \nabla \cdot \left( \rho \bfv \right) = 0 , \label{eq:eom_1} \\
& \pa_t \bfv + \left( \bfv \cdot \nabla \right) \bfv  = - \frac{\nabla P}{\rho} -\nabla \Phi - 2 \bfOmegap \times \bfv - \bfOmegap \times \left( \bfOmegap \times \bfR \right) , \label{eq:eom_2}
\end{align}
where $\bfv$ is the velocity in the rotating frame, $\rho$ is the surface density, $P$ is the pressure, $\Phi$ is an external gravitational potential, $- 2 \bfOmegap \times \bfv$ is the Coriolis force, $ - \bfOmegap \times \left( \bfOmegap \times \bfR \right)$ is the centrifugal force. The explicit form of the potential $\Phi$ will be specified later.

\subsection{Spiral coordinates}

\begin{figure}
\includegraphics[width=\columnwidth]{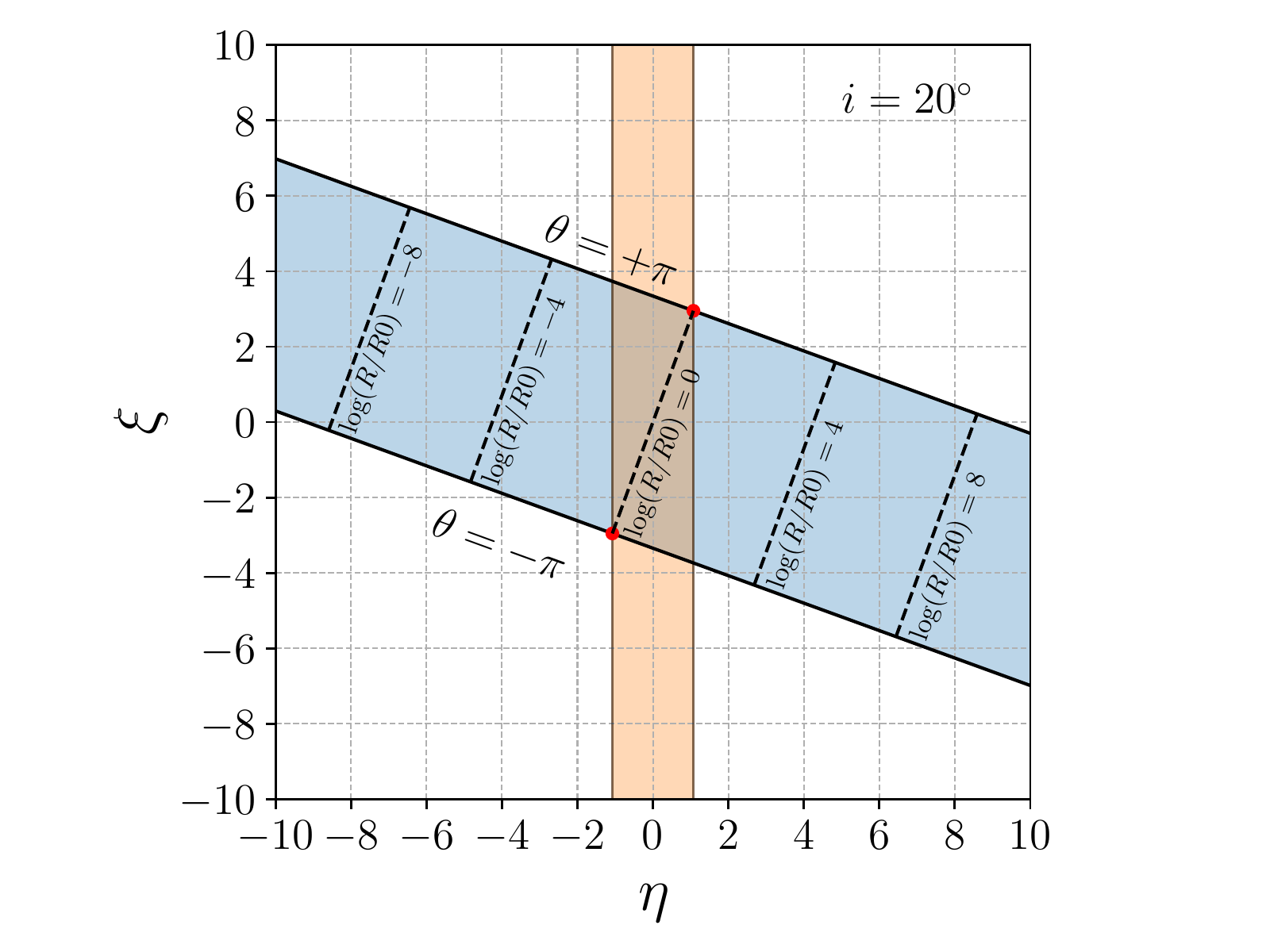}
\caption{\emph{Blue shaded:} domain of the $(\eta,\xi)$ coordinates if we require that $\theta \in [ -\pi, \pi]$. \emph{Orange shaded:} another possible choice for the domain. This choice is more natural in terms of the $(\eta,\xi)$ coordinates, but corresponds to a more convoluted $\theta$ domain. The two vertical lines that bound the shaded orange region correspond to $\eta=\pm \pi \sin(i)$. The two red dots represent the same \emph{physical} point and correspond to the single red dot in Figure \ref{fig:spiralcoord2}.}
\label{fig:spiralcoord1} 
\end{figure}

\begin{figure}
\includegraphics[width=\columnwidth]{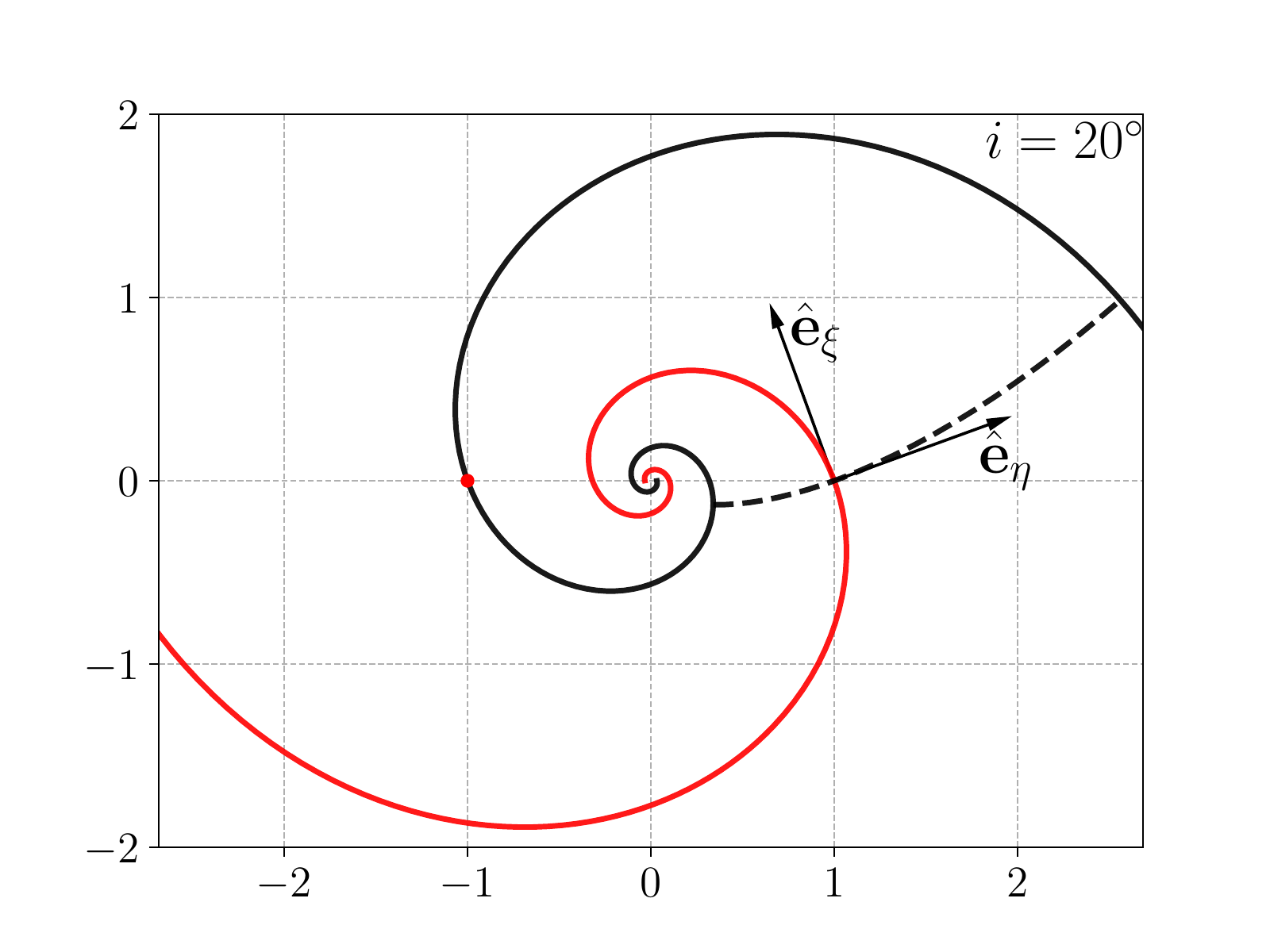}
\caption{The spiral coordinate system for $i=20^\circ$ and $R_0=1$. \emph{Black solid:} the line $\eta=\pm \pi\sin(i)={\rm constant}$ and $\xi=[0,\infty]$. This corresponds to either of the two vertical shaded lines that bound the orange domain in Figure \ref{fig:spiralcoord1}. \emph{Red solid:} the line $\eta=0={\rm constant}$ and $\xi=[0,\infty]$. This corresponds to the vertical line passing through the origin in Figure \ref{fig:spiralcoord1}. \emph{Black dashed:} the line $\eta=[-\pi\sin(i),\pi\sin(i)]$ and $\xi=0 = {\rm constant}$. The single red dot corresponds to both red dots in Figure \ref{fig:spiralcoord1}. Note that the point $\{\eta=0,\xi=0\}$ corresponds to the point $\{R=1,\theta=0\}$.}
\label{fig:spiralcoord2} 
\end{figure}

Following \citet{Roberts1969}, we define the following spiral coordinates:
\begin{align}
\eta & = \log\left(R/R_0\right) \cos(i)+\theta\sin(i), \label{eq:spiralcoord1} \\
\xi   & = -\log\left(R/R_0\right) \sin(i)+\theta\cos(i). \label{eq:spiralcoord2}
\end{align}
The inverse relations are:
\begin{align}
\log\left(R/R_0\right) = \eta \cos(i) - \xi \sin(i), \label{eq:coord3} \\
\theta = \eta \sin(i)+ \xi \cos(i), \label{eq:coord4}
\end{align}
where $R$, $\theta$ are usual polar coordinates and $R_0$ and $i$ are constants. There are several possible choices for the domain of the $(\eta,\xi)$ coordinates, two of which are shown in Figure \ref{fig:spiralcoord1}. In the following, we use the domain corresponding to the orange shaded region in Figure \ref{fig:spiralcoord1}:
\begin{equation}
\eta = [-\pi \sin(i), \pi \sin(i)]\,,		\qquad \xi = [0,\infty] \,.
\end{equation}
Figure \ref{fig:spiralcoord2} shows lines of constant $\eta$ and $\xi$ for this choice of the domain and $R_0=1$, $i=20\degree$. The origin $(\eta=0,\xi=0)$ corresponds to the point $(R=1,\theta=0)$. As we increase the coordinate $\eta$ at constant $\xi$, we move perpendicularly to the spirals. Note that when we cross the value $\eta=\pi \sin(i)$ (reappearing on the other side at $\eta=-\pi \sin(i))$, there is a jump in the coordinate $\xi$ of $\Delta \xi = 2\pi \cos(i)$. Therefore, when working with these coordinates all physical quantities such as for example the density must satisfy the following condition:
\begin{equation} \label{eq:boundary1}
\rho(\xi,\eta) = \rho(\eta+2\pi\sin(i),\xi+2\pi\cos(i)).
\end{equation}
The unit vectors in spiral coordinates are:
\begin{align}
\hat{e}_{\eta} & = \cos(i) \hat{e}_{R} + \sin(i) \hat{e}_{\theta}, \\ 
\hat{e}_{\xi}   & = - \sin(i) \hat{e}_{R} + \cos(i) \hat{e}_{\theta}.
\end{align}
Straightforward calculations show that the gradient in spiral coordinates is:
\begin{equation}
{\nabla}  =  \frac{1}{R}\left( \hat{e}_{\eta} \de{}{\eta} + \hat{e}_{\xi} \de{}{\xi} \right)
\end{equation}
and the derivatives of the unit vectors are:
\begin{align}
\de{\hat{e}_{\eta}}{\eta} & = \sin(i) \hat{e}_{\xi}, \qquad  & \de{\hat{e}_{\eta}}{\xi} &= \cos(i) \hat{e}_{\xi}, \\
\de{\hat{e}_{\xi}}{\eta}   & = -\sin(i) \hat{e}_{\eta}, \qquad & \de{\hat{e}_{\xi}}{\xi} &= - \cos(i) \hat{e}_{\eta} .
\end{align}

\subsection{Equations of motion in spiral coordinates}
Using the relations of the previous subsection it is straightforward to rewrite Equations \eqref{eq:eom_1} and \eqref{eq:eom_2} in spiral coordinates. The continuity equation becomes:
\begin{equation} 
\pa_t \rho + \frac{1}{R}\left[ \pa_{\eta} \left(\rho v_{\eta}\right) + \pa_{\xi} \left(\rho v_{\xi}\right) + \rho \left(v_{\eta} \cos(i) - v_{\xi} \sin(i) \right)\right] = 0,
\label{eq:continuity}
\end{equation}
and the Euler equation:
\begin{align}
	\pa_t v_{\eta} + &\frac{1}{R} \left[ v_{\eta} \left(\pa_{\eta} v_{\eta}\right) + v_{\xi} \left(\pa_{\xi} v_{\eta}\right) - v_{\xi} v_{\eta} \sin(i) - v_{\xi}^2 \cos(i) \right] \nonumber \\
		& = - \frac{1}{R} \frac{ \pa_{\eta} P}{\rho} - \frac{1}{R} \pa_{\eta} \Phi + 2 \Omegap v_{\xi} +\cos(i) \Omegap^2, \label{eq:eulerspiral1} \\
	\pa_t v_{\xi}   + &\frac{1}{R} \left[ v_{\eta} \left(\pa_{\eta} v_{\xi}\right) + v_{\xi} \left(\pa_{\xi} v_{\xi}\right) + v_{\eta} v_{\xi} \cos(i) + v_{\eta}^2 \sin(i) \right] \nonumber \\ 
		& = - \frac{1}{R} \frac{ \pa_{\xi} P}{\rho} - \frac{1}{R} \pa_{\xi} \Phi - 2\Omegap v_{\eta} -\sin(i) \Omegap^2 . \label{eq:eulerspiral2}
\end{align}

\subsection{Split into circular and spiral components}
We now write
\begin{equation}
\begin{split} \label{eq:split1}
& \bfv = \bfv_{\rm c} + \bfv_{\rm s}\\
& \rho = \rho_{\rm c} + \rho_{\rm s}\\
& P = P_{\rm c} + P_{\rm s} \\
& \Phi = \Phi_{\rm c} + \Phi_{\rm s}
\end{split}
\end{equation}
where the subscript $\rm c$ refers to a steady state axisymmetric solution in the axisymmetric potential $\Phi_{\rm c}$ and $\rm s$ to a ``spiral'' departure from the axisymmetric solution. We assume that the axisymmetric solution is of the type
\begin{align}
\rho_{\rm c} & = {\rm constant} \\
P_{\rm c} & = {\rm constant} \\
\bfv_{\rm c} & = \left( \bfOmega(R) - \bfOmegap \right) \times \bfR \label{eq:vc}
\end{align}
where $\bfOmega = \Omega \hatez$. In spiral coordinates we can write:
\begin{equation} \label{eq:R}
\bfR = R \left( \cos(i) \hat{e}_{\eta} - \sin(i) \hat{e}_{\xi} \right).
\end{equation}
The circular velocity can therefore be written as:
\begin{align}
v_{{\rm c} \eta} & = \left(\Omega(R) - \Omegap\right) R \sin(i) \,,\\
v_{{\rm c} \xi} & = \left(\Omega(R) - \Omegap\right) R \cos(i)\,.\label{eq:vcircxi}
\end{align}

\subsection{Radial spacing between spiral arms}
Starting from the spiral arm defined by the relation $\eta=-\pi \sin(i)$ (black line in Figure \ref{fig:spiralcoord2}) and moving along the coordinate $\eta$ at constant $\xi$, we meet the next arm after $\Delta \eta = 2 \pi \sin(i) / m$ (and $\Delta \xi=0$ by definition). Using these in Equations \eqref{eq:coord3}-\eqref{eq:coord4}, we find the corresponding change in $R$ and $\theta$ as we move from one arm to the next:
\begin{align} 
 \log\left(1 + \frac{L}{R}\right)& = \Delta\eta \cos(i) - \Delta\xi \sin(i)  = \frac{2 \pi \sin(i) \cos(i)}{m}, \\
 \Delta\theta &= \Delta\eta \sin(i)+ \Delta\xi \cos(i)  =  \frac{\pi \sin(i)^2}{m}, 
\end{align}
where $L$ is defined as the radial separation between two arms. In the limit $\sin(i) \ll 1$ these become
\begin{align} 
\frac{L}{R}& \simeq \frac{2 \pi\sin(i)}{m}, \label{eq:L} \\ 
 \Delta\theta &  \simeq 0.
\end{align}

\subsection{Approximations} \label{sec:approximations}
Following \citet{Roberts1969} (see also \citealt{Balbus1988}) we now approximate the equations of motion under the following assumptions:
\begin{enumerate}
\item The pitch angle is small,
\begin{equation} \sin i \ll 1. \end{equation}
\item The velocity $\vcxi \sim \Omega R$ is much greater than $v_{{\rm s}\xi}$, $v_{{\rm s}\eta}$ and $\vceta\sim\Omega R\sin(i)$. The latter are all comparable in size. Thus
\begin{equation} v_{{\rm c} \xi} \sim  \Omega R \gg v_{{\rm c}\eta} \sim v_{{\rm s}\eta} \sim v_{{\rm s}\xi} \sim \Omega R \sin(i) \,. \end{equation}
\item All quantities with subscript `s' vary much faster in the direction $\hat{e}_\eta$ (with a length-scale $L$, corresponding to $\Delta \eta \simeq 2\pi \sin(i)/m$), than in the direction $\hat{e}_\xi$ (length-scale $R$, corresponding to $\Delta \xi \simeq 2\pi \cos(i)$). Thus
\begin{equation} \pa_\eta \sim \frac{1}{\sin(i)}, \quad \pa_\xi \sim 1. \qquad \text{(spiral quantities)} \end{equation}  
\item All quantities with subscript 'c' vary with a length-scale $R$ in the direction $R$ (in the other directions they do not vary since they are axisymmetric by definition). According to Equations \eqref{eq:spiralcoord1} and \eqref{eq:spiralcoord2}, in order to double $R$ we need to move by $\Delta \xi=\log(2)/\sin(i)$ in the $\xi$ direction (at constant $\eta$) or by $\Delta\eta=\log(2)/\cos(i)$ in the $\eta$ direction (at constant $\xi$). Thus
\begin{equation} \pa_\eta \sim 1, \quad \pa_\xi \sim \sin(i). \qquad \text{(background quantities)} \end{equation}  
\item The sound speed of the gas is smaller or comparable to the spiral velocities:
\begin{equation}
c_{\rm s} \lesssim \Omega R \sin(i) \,. 
\end{equation}
\item The strength of the spiral potential is comparable to the spiral velocities squared:
\begin{equation}
\Phi_{\rm s} \sim \left[\Omega R \sin(i)\right]^2 \,.
\end{equation}
\end{enumerate}
As a consequence of the first assumption, we have that the radial spacing between the spiral arms $L$ is much smaller than $R$ (see Equation \ref{eq:L}):
\begin{equation} \label{eq:L2} \frac{L}{R} \simeq \frac{2 \pi \sin(i)}{m} \ll 1 \,. \end{equation} 

\subsection{Local coordinates}

We define local $(x,y)$ coordinates centred on a small patch around $R_0$ as follows:
\begin{align}
x & = R_0 \eta \,,\\
y & = R_0 \xi \,.
\end{align}
The point $(x=0,y=0)$ corresponds to the point $(R=R_0,\theta=0)$. We require the range of $\eta$ to be from one spiral arm to the next, so $\eta = [-\pi \sin(i)/m, \pi \sin(i)/m]$, and the range of $\xi$ to be comparable. Thus we have 
\begin{equation}
|x| \lesssim R_0 \sin(i) \ll R_0 \qquad |y| \lesssim R_0 \sin(i) \ll R_0  \label{eq:xapprox}
\end{equation}
Under these approximations, we can rewrite the the radius as (see Equation \ref{eq:coord3}):
\begin{align}
R & = R_0 \exp\left( \eta \cos(i) - \xi \sin(i) \right) \,, \\
 & \simeq  R_0 \left(1 + \eta \right)\,, \label{eq:Rapproxi} \\
 & = R_0 + x\,. \label{eq:Rapprox}
\end{align}
The derivatives with respect to the $(x,y)$ coordinates are
\begin{equation}
\pa_x = \frac{1}{R_0} \pa_\eta\,, \quad \qquad \pa_y = \frac{1}{R_0} \pa_\xi \,. \label{eq:xder}
\end{equation}
The velocities in $(x,y)$ coordinates are the same as in spiral coordinates:
\begin{equation}
v_x = v_\eta \,, \quad \qquad v_y = v_\xi \,. \label{eq:vxy}
\end{equation}

\subsection{Approximating the continuity equation}
We now want to approximate the continuity equation \eqref{eq:continuity} to leading order in $\sin(i)$. To do this, we first substitute \eqref{eq:split1} into \eqref{eq:continuity} and eliminate all the terms that simplify because the axisymmetric solution (subscript `c') also separately satisfies \eqref{eq:continuity}.  This gives:
\begin{align} 
	\pa_t \rho_{\rm s} & + \frac{1}{R}\left[ \pa_{\eta} \left(\rho_{\rm c} v_{{\rm s} \eta }\right) + \pa_{\xi} \left(\rho_{\rm c} v_{{\rm s}\xi}\right) + \rho_{\rm c} \left(v_{{\rm s}\eta} \cos(i) - v_{{\rm s}\xi} \sin(i) \right)\right] \nonumber \\
				    & + \frac{1}{R}\left[ \pa_{\eta} \left(\rho_{\rm s} v_{{\rm c} \eta }\right) + \pa_{\xi} \left(\rho_{\rm s} v_{{\rm c}\xi}\right) + \rho_{\rm s} \left(v_{{\rm c}\eta} \cos(i) - v_{{\rm c}\xi} \sin(i) \right)\right] \nonumber \\
				    & + \frac{1}{R}\left[ \pa_{\eta} \left(\rho_{\rm s} v_{{\rm s} \eta }\right) + \pa_{\xi} \left(\rho_{\rm s} v_{{\rm s}\xi}\right) + \rho_{\rm s} \left(v_{{\rm s}\eta} \cos(i) - v_{{\rm s}\xi} \sin(i) \right)\right] \nonumber \\
				    & = 0
\label{eq:continuity2}
\end{align}
We now have to estimate the order of each term according to the relations listed in Section \ref{sec:approximations} and eliminate the negligible ones. For the terms in the first row of Equation \eqref{eq:continuity2} we have:
\begin{align}  
\pa_\eta \left( \rhoc \vseta\right) & \sim \frac{1}{\sin(i)} \rhoc \Omega R \sin(i) \sim \rhoc \Omega R \\
\pa_\xi \left( \rhoc \vsxi \right) & \sim \rhoc \Omega R \sin(i) \\
\rhoc \vseta \cos(i) & \sim \rhoc \Omega R \sin(i) \\
\rhoc \vseta \sin(i) & \sim \rhoc \Omega R [\sin(i)]^2
 \end{align}
To leading order in $\sin(i)$ we only need to keep $\pa_\eta \left( \rhoc \vseta\right)$ (all the others are negligible compared to this). For the terms in the second row we have (remember $\pa_\eta\rhos\sim \rhos/\sin(i)$, while $\pa_\eta \vceta \sim \vceta$):
 \begin{align}  
\pa_\eta \left( \rhos \vceta\right) & =   \vceta \pa_\eta(\rhos) + \rhos \pa_\eta(\vceta) \nonumber \\
						&  \sim \rhos \Omega R + \rhos \Omega R \sin(i) \\
\pa_\xi \left( \rhos \vcxi \right) & = \rhos \pa_\xi\left(\vcxi \right) + \vcxi \pa_\xi \left( \rhos \right) \\
					     & \sim \rhos \Omega R \sin(i) +  \rhos \Omega R \\
\rhos \vceta \cos(i) & \sim \rhos \Omega R \sin(i) \\
\rhos \vcxi \sin(i) & \sim \rhos \Omega R \sin(i)
 \end{align}
Thus we need to keep $\vceta \pa_\eta(\rhos)$ and $\vcxi \pa_\xi \left( \rhos \right)$. Proceeding similarly for the third row and approximating $1/R \simeq 1/R_0$ everywhere in Equation \eqref{eq:continuity2} (which is correct to leading order in $\sin(i)$, see Equations \ref{eq:Rapprox} and \ref{eq:xapprox}) we find that Equation \eqref{eq:continuity2} reduces to:
\begin{align}
\pa_t \rho_s & + \frac{1}{R_0} \left[ \pa_\eta \left(\rhoc \vseta\right) + \pa_\eta \left( \rhos \vseta \right) + \vceta \pa_\eta(\rhos) + \vcxi \pa_\xi (\rhos) \right] \nonumber \\
		  & + O(\sin(i)) = 0 \,. \label{eq:continuity3}
\end{align}
This is the minimal amount of terms that we need to keep. We can however some terms of order $\sin(i)$ to make the final equation appear more familiar while committing a negligible error of $O(\sin(i))$. Going back to Equation \eqref{eq:continuity}, we see that all the terms of order $O(1)$ that appear in Equation \eqref{eq:continuity3} originate from the first two terms inside the square parentheses. Therefore, we can approximate Equation \eqref{eq:continuity} as:
 \begin{equation} 
	\pa_t \rho + \frac{1}{R_0}\left[ \pa_{\eta} \left(\rho v_{\eta}\right) + \pa_{\xi} \left(\rho v_{\xi}\right) \right] + O(\sin(i)) = 0,
\label{eq:continuity4}
\end{equation}
This equation is equivalent to \eqref{eq:continuity3} to order $O(\sin(i))$, but is more useful because it looks like the normal continuity equation. Finally, we can use \eqref{eq:xder} to re-express \eqref{eq:continuity4} in local coordinates as:
 \begin{equation} 
\boxed{	\pa_t \rho +  \pa_{x} \left(\rho v_{x}\right) + \pa_{y} \left(\rho v_{y}\right)  + O(\sin(i)) = 0 }
\label{eq:continuity5}
\end{equation}
This equation coincides with Equation (2.1c) of \citet{Balbus1988} and with Equation (2) of \citet{Kim32014}.

\subsection{Approximating the Euler equation} \label{sec:approxeuler}
We split the Euler equation \eqref{eq:eulerspiral1} and \eqref{eq:eulerspiral2} into a `circular' and `spiral' component and then approximate to leading order in $\sin(i)$. Substituting \eqref{eq:split1} into \eqref{eq:eulerspiral1} and \eqref{eq:eulerspiral2}, and eliminating all the terms that simplify because the axisymmetric solution also separately satisfies Equation \eqref{eq:split1}, we obtain respectively:
\begin{align}
	\pa_t \vseta 
	+ &\frac{1}{R} \left[ \vseta \left(\pa_{\eta} \vseta \right) + \vsxi \left(\pa_{\xi} \vseta \right) - \vsxi \vseta \sin(i) - \vsxi^2 \cos(i) \right] \nonumber \\
	+ &\frac{1}{R} \left[ \vseta \left(\pa_{\eta} \vceta \right) + \vsxi \left(\pa_{\xi} \vceta \right) - \vsxi \vceta \sin(i) - \vsxi\vcxi \cos(i) \right] \nonumber \\
	+ &\frac{1}{R} \left[ \vceta \left(\pa_{\eta} \vseta \right) + \vcxi \left(\pa_{\xi} \vseta \right) - \vcxi \vseta \sin(i) - \vcxi\vsxi \cos(i) \right] \nonumber \\
		& = - \frac{1}{R} \frac{ \pa_{\eta} P_{\rm s}}{\rho} - \frac{1}{R} \pa_{\eta} \Phi_{\rm s} + 2 \Omegap \vsxi\,. \label{eq:eulerspiral5}
\end{align}
and
\begin{align}
	\pa_t \vsxi   
	+ &\frac{1}{R} \left[ \vseta \left(\pa_{\eta} \vsxi \right) + \vsxi \left(\pa_{\xi} \vsxi \right) + \vseta \vsxi \cos(i) + \vseta^2 \sin(i) \right] \nonumber \\
	+ &\frac{1}{R} \left[ \vseta \left(\pa_{\eta} \vcxi \right) + \vsxi \left(\pa_{\xi} \vcxi \right) + \vseta \vcxi \cos(i) + \vseta\vceta \sin(i) \right] \nonumber \\
	+ &\frac{1}{R} \left[ \vceta \left(\pa_{\eta} \vsxi \right) + \vcxi \left(\pa_{\xi} \vsxi \right) + \vceta \vsxi \cos(i) + \vceta\vseta \sin(i) \right] \nonumber \\ 
		& = - \frac{1}{R} \frac{ \pa_{\xi} P_{\rm s}}{\rho} - \frac{1}{R} \pa_{\xi} \Phi_{\rm s} - 2\Omegap \vseta \,. \label{eq:eulerspiral6}
\end{align}
So far we have not performed any approximation. We now estimate the order of each term in these equations using the relations in Section \ref{sec:approximations}. We want to keep only the leading order in $\sin(i)$. For the terms within the square parentheses in Equation \eqref{eq:eulerspiral5} we have:
\begin{align}
\vseta \left(\pa_\eta \vseta \right) & \sim [\Omega R]^2 \sin(i) \qquad (*) \label{eq:keep1} \\
\vsxi \left( \pa_\xi \vseta \right) & \sim [\Omega R]^2 [\sin(i)]^2 \\
\vsxi \vseta \sin(i) & \sim [\Omega R]^2 [\sin(i)]^3 \\
\vsxi^2 \cos(i) & \sim [\Omega R]^2 [\sin(i)]^2 \\
\vseta \left( \pa_\eta \vceta\right) & \sim [\Omega R]^2 [\sin(i)]^2 \\
\vsxi \left( \pa_\xi \vceta \right) &\sim [\Omega R]^2  [\sin(i)]^3  \\
\vsxi \vceta \sin(i) & \sim [\Omega R]^2 [\sin(i)]^3 \\
\vsxi \vcxi \cos(i) & \sim [\Omega R]^2 \sin(i)  \qquad (*) \label{eq:keep2} \\
\vceta \left( \pa_\eta \vseta \right)  & \sim [\Omega R]^2 \sin(i) \qquad (*) \label{eq:keep3} \\
\vcxi \left( \pa_\xi \vseta \right) & \sim [\Omega R]^2 \sin(i)  \qquad (*)\label{eq:keep4} \\
\vcxi \vseta \sin(i) & \sim [\Omega R]^2 [\sin(i)]^2 \\
\vcxi \vsxi \cos(i) & \sim [\Omega R]^2 \sin(i)  \qquad (*) \label{eq:keep5}
\end{align}
Thus to leading order in $\sin(i)$ we need to keep terms marked with $(*)$. To the same order we can also approximate $1/R \simeq 1/R_0$ everywhere in Equation \eqref{eq:eulerspiral5} (see Equations \ref{eq:Rapprox} and \ref{eq:xapprox}). Using the relations given in Section \ref{sec:approximations} we see that all the terms on the right-hand-side of Equation \eqref{eq:eulerspiral5} are of order $[\Omega R]^2 \sin(i)$, so we have to keep them. We can put $\cos(i)\simeq 1$ in terms \eqref{eq:keep2} and \eqref{eq:keep5}. To the same order we can also approximate $\vcxi \simeq (\Omega_0 -\Omegap) R_0 = {\rm constant}$ (see Equation \ref{eq:vcircxi}) in terms \eqref{eq:keep2}, \eqref{eq:keep4} and \eqref{eq:keep5}, where $\Omega_0=\Omega(R_0)$. Putting everything together, we can rewrite Equation \eqref{eq:eulerspiral5} as:
\begin{align}
	\pa_t \vseta 
	+ &\frac{1}{R_0} \left[ \left( \vseta + \vceta \right) \left(\pa_{\eta} \vseta \right) + \vcxi \left(\pa_{\xi} \vseta \right)  \right] \nonumber \\
		& = - \frac{1}{R} \frac{ \pa_{\eta} P_{\rm s}}{\rho} - \frac{1}{R} \pa_{\eta} \Phi_{\rm s} + 2 \Omega_0 \vsxi  + O\left([\sin(i)]^2\right) \,.
\end{align}
Committing a negligible error of order $O(\sin(i))$ we can add a term $\vsxi \left(\pa_{\xi} \vseta \right)$ inside the square parentheses, to make the result look more similar to the usual Euler equation in Cartesian coordinates. Using \eqref{eq:split1}, \eqref{eq:xder} and \eqref{eq:vxy} we can finally rewrite Equation \eqref{eq:eulerspiral5} as:
\begin{subequations}
\begin{empheq}[box=\widefbox]{align}
	\pa_t v_{{\rm s}x} 
	+ &  v_x \left(\pa_x v_{{\rm s}x} \right) + v_y \left(\pa_y v_{{\rm s}x} \right)   \nonumber \\
		& = - \frac{ \pa_x P_{\rm s}}{\rho} - \pa_x \Phi_{\rm s} + 2 \Omega_0 v_{{\rm s}y}  + O\left([\sin(i)]^2\right) 
\end{empheq} \label{eq:finaleuler1}
\end{subequations}
Now we repeat similar calculations for Equation \eqref{eq:eulerspiral6}. For the various terms within the square parentheses in this equation we have:
\begin{align}
\vseta \left( \pa_\eta \vsxi \right) & \sim [\Omega R]^2 \sin(i) \qquad (*) \label{eq:1keep} \\
\vsxi \left( \pa_\xi \vsxi \right) & \sim [\Omega R]^2 [\sin(i)]^2 \\
\vseta\vsxi \cos(i) &  \sim [\Omega R]^2 [\sin(i)]^2\\
\vseta^2 \sin(i) & \sim  [\Omega R]^2 [\sin(i)]^3 \\
\vseta \left(\pa_\eta \vcxi \right) & \sim [\Omega R]^2 \sin(i) \qquad (*) \label{eq:2keep} \\
\vsxi \left( \pa_\xi \vcxi \right) & \sim [\Omega R]^2 [\sin(i)]^2 \label{eq:1keepextra} \\
\vseta \vcxi \cos(i) & \sim [\Omega R]^2 \sin(i)  \qquad (*) \label{eq:3keep} \\
\vseta\vceta \sin(i) & \sim [\Omega R]^2 [\sin(i)]^3 \\
\vceta \left( \pa_\eta \vsxi \right) & \sim [\Omega R]^2 \sin(i)  \qquad (*) \label{eq:4keep} \\
\vcxi \left(\pa_\xi \vsxi \right) & \sim [\Omega R]^2 [\sin(i)]^2 \label{eq:2keepextra} \\
\vceta \vsxi \cos(i) & \sim [\Omega R]^2 [\sin(i)]^2 \\
\vceta \vseta \sin(i) & \sim [\Omega R]^2 [\sin(i)]^3
\end{align}
To leading order in $\sin(i)$ we need to keep terms that are marked with $(*)$. To the same order the derivative in term \eqref{eq:2keep} can be rewritten as (see Equations \ref{eq:vcircxi} and \ref{eq:Rapproxi}):
\begin{align}
\pa_\eta \vcxi & \simeq \left( \Omega_0 - \Omegap \right) R_0 + \left( \frac{\di\Omega}{\di R}\right)_{|R_0} R_0^2 \,, \\
						& = \left( \Omega_0 - \Omegap \right) R_0 - q  \Omega_0 R_0 \,,
\end{align}
where we introduced the shear parameter
\begin{equation}
q = - \left( \frac{\di \log(\Omega)}{\di \log(R)} \right)_{|R_0} \,.
\end{equation}
To the same order term \eqref{eq:3keep} can be written as:
\begin{equation}
\vseta \vcxi \cos(i) \simeq \vseta ( \Omega_0 - \Omegap ) R_0\,.
\end{equation}
To the same order we can put $\cos(i)=1$ in term \eqref{eq:3keep} and approximate $1/R\simeq1/R_0$ everywhere in Equation \eqref{eq:eulerspiral6}. The terms with $P_{\rm s}$ and $\Phi_{\rm s}$ on the RHS of Equation \eqref{eq:eulerspiral6} are of order $[\Omega R]^2 [\sin(i)]^2$ and could be neglected, but we keep them, committing a negligible error of order $O([\sin(i)]^2)$. We also keep terms \eqref{eq:1keepextra} and \eqref{eq:2keepextra} committing negligible errors. Putting everything together and using \eqref{eq:split1}, \eqref{eq:xder} and \eqref{eq:vxy} we find:
\begin{subequations}
\begin{empheq}[box=\widefbox]{align}
	\pa_t v_{{\rm s}y}   
	+ & v_x \left(\pa_x v_{{\rm s}y} \right) + v_y \left(\pa_y v_{{\rm s} y} \right) \nonumber \\ 
		& = - \frac{ \pa_y P_{\rm s}}{\rho} - \pa_y \Phi_{\rm s} - 2\Omega_0 v_{{\rm s}x} + q \Omega_0 v_{{\rm s}x} + O\left([\sin(i)]^2\right) 
\end{empheq} \label{eq:finaleuler2}
\end{subequations}
Equations \eqref{eq:finaleuler1} and \eqref{eq:finaleuler2} agree with Equations (2.1a) and (2.1b) of \citet{Balbus1988} and with Equation (3) of \citet{Kim32014}. Some remarks:
\begin{enumerate} 
\item Despite velocities $v_x$ and $v_y$ being in the frame rotating at $\Omegap$, the Coriolis term that appears in Equations \eqref{eq:finaleuler1} and \eqref{eq:finaleuler2} is calculated using $\Omega_0$, \emph{not} $\Omega_{\rm p}$.
\item In deriving Equations \eqref{eq:continuity5}, \eqref{eq:finaleuler1}, and \eqref{eq:finaleuler2}, we have \emph{not} expanded to first order in the quantities with subscript 's', as we would have done in a standard linear analysis. Indeed, we have kept quadratic terms such as $\vsx (\pa_x \vsx)$, which we would not have kept in a linear analysis. Instead, the small parameter in the present expansion is $\sin(i)$.
\item In solving Equations \eqref{eq:continuity5}, \eqref{eq:finaleuler1}, and \eqref{eq:finaleuler2}, the circular velocity $\bfv_{\rm c}$ must be specified, because it enters through the terms $v_x = v_{{\rm c} x} + v_{{\rm s} x} $ and $v_y = v_{{\rm c} y} +  v_{{\rm s} y}$. However, in all instances in which $\bfv_{\rm c}$ appears, it can be considered a constant. This can be shown by considering one by one the various terms that contain it. For example, in Equation \eqref{eq:finaleuler1} we have the term 
\begin{align}
v_y (\pa_y\vsx) & = \vcy (\pa_y \vsx) + \vsy (\pa_y \vsx)\,. \label{eq:vapprox}
\end{align}
The circular velocity can be expanded as (see Equations \ref{eq:vcircxi}):
\begin{align}
v_{{\rm c} x} & = \vcxo + O\left([\sin(i)]^2\right)  \label{eq:vcircx} \,,\\
v_{{\rm c} y} & = \vcyo + O\left([\sin(i)]\right) \,.\label{eq:vcircy}
\end{align}
where
\begin{align}
 \vcxo & = \left(\Omega_0 - \Omegap\right) R_0 \sin(i)  \label{eq:vcircx0} \,,\\
\vcyo  & = \left(\Omega_0 - \Omegap\right) R_0 \label{eq:vcircy0}\,,
\end{align}
are the circular velocities at $R=R_0$. 
Since according to the relations in Section \ref{sec:approximations} we have $(\pa_y \vsx) = (1/R_0) \pa_\xi ( \vsxi) \sim \Omega R \sin(i) $, when we substitute \eqref{eq:vcircy} into the first term on the right hand side of \eqref{eq:vapprox} we obtain:
\begin{align}
\vcy (\pa_y \vsx) & = \vcyo (\pa_y \vsx) + O([\sin(i)]^2) \,. \label{eq:vapprox_rhs}
\end{align}
When we put this back into \eqref{eq:vapprox} and then into \eqref{eq:finaleuler1}, we see that we can approximate $\vcy \simeq \vcyo$ to the same level of approximation under which \eqref{eq:finaleuler1} is valid, i.e. $O([\sin(i)]^2)$. Repeating the same argument with all the terms in which $\vcx$ and $\vcy$ appear in Equations \eqref{eq:continuity5}, \eqref{eq:finaleuler1}, and \eqref{eq:finaleuler2}, one sees that we can approximate everywhere $\vcx \simeq \vcxo$ and $\vcy \simeq \vcyo$.
\end{enumerate}

\subsection{Final equations}\label{sec:finaleq}
We now re-express the equations in the form used in the main text. Equation \eqref{eq:continuity5} is already in the same form as \eqref{eq:2e2}. To bring Equations \eqref{eq:finaleuler1} and \eqref{eq:finaleuler2} in the form \eqref{eq:2e1}, note that as specified in item (iii) in Section \ref{sec:approxeuler} above we can consider $v_{\rm c}$ to be a constant everywhere in these equations. So we can replace $v_{\rm{s}}$ with $v$ in all terms containing a derivative, and we can substitute $\vsx = v_x - \vcxo$ and $\vsy = v_y - \vcyo$ (see Equations~\ref{eq:vcircx0} and \ref{eq:vcircy0}). We arrive at:
\begin{equation}
\pa_t\mv+\left(\mv\cdot\nabla\right)\mv=q\Omega_0 v_x \hatey-\frac{\nabla P}{\rho}-\nabla\Phis -2\bfOmega_0\times\mv+\bfG,\label{2.60}
\end{equation}
where $\bfG$ is a constant given by:
\begin{equation}
\bfG= 2\Omega_0 (\Omega_0 - \Omegap)R_0 \left( \sin(i)\left(1 - \frac{q}{2}\right)\hatey-\hatex \right).\label{2.61}
\end{equation}
Now we perform the following Galileian transformation to put ourselves in a frame that moves along the spiral arm with a speed equal to the circular velocity.:
\begin{align}
\begin{split}\label{9.1}
x'=x,\;\;y'=y - \vcyo t,\;\;t'=t,
\end{split}
\end{align}
Equation~\eqref{eq:continuity5} is invariant under this transformation and remains unchanged. All terms in Equation~\eqref{2.60} are invariant except the last two, which substituting $v_y = v_y' + \vco$ become:
\begin{align}
\begin{split}
-2\Omega_0\times\mv+\bfG &= -2\Omega\times\mv' + 2\Omega_0\vco \sin(i)\left(1 - \frac{q}{2}\right)\hatey.
\end{split}
\end{align}
Therefore the final Euler equation becomes:
\begin{equation} \label{eq:finaleuler3}
\boxed{
\pa_t\mv+\left(\mv\cdot\nabla\right)\mv=q\Omega_0 v_x \hatey-\frac{\nabla P}{\rho}-\nabla\Phis -2\bfOmega_0\times\mv+F \left(1 - \frac{q}{2}\right) \hatey }
\end{equation}
where we have dropped the primes for simplicity and $F$ is a constant given by:
\begin{equation}
F= 2\Omega_0 (\Omega_0 - \Omegap)R_0 \sin(i).
\end{equation}
Equation \eqref{eq:finaleuler3} coincides with Equation~\eqref{eq:2e1}.

\section{Numerical solution of the steady state equations}\label{sec:steady}
Here we describe the numerical procedure used to solve Equations~\eqref{eq:2A1} and \eqref{eq:2A2}. We use the `shooting method' \citep[e.g.][]{Press2007}. Our procedure is similar to those of \citet{Shu1973} and \citet{Kim32014}. 

As mentioned in the main text, we are interested in solutions that contain a shock. These solutions must  contain a sonic point $\xs$, defined as the point where $v_x(\xs) = \cs$. We do not know the position of the sonic point a priori, so we take an initial guess for the position of the sonic point. We then determine the velocity $v_y(\xs)=\Phis'(\xs)/(2\Omega_0)$ at the sonic point by using the fact that Equation~\eqref{eq:2A1} should not be singular at the sonic point.

\begin{figure}
\centering
\includegraphics[width=1\linewidth]{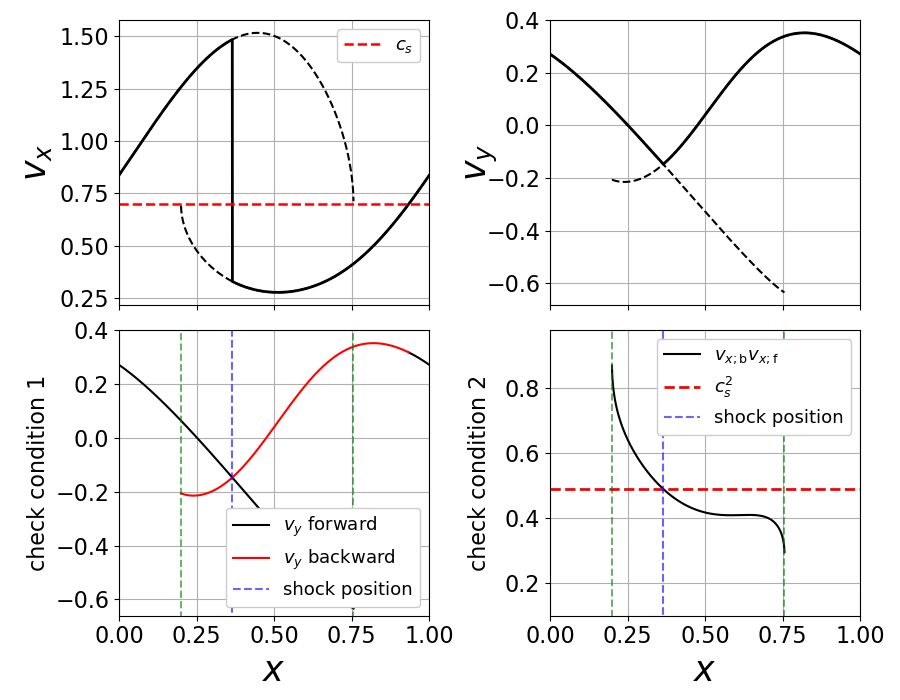}
\caption{An example numerical solution to Equations \eqref{eq:2A1} and \eqref{eq:2A2}. These steady state solutions are used as initial conditions for our simulations. For this example, $q=0$, $L_x=1$, $\cs=0.7$ and $\Phi_0=0.25$. The top panels show the final solution as solid lines, and the discarded continuations of the forward and backward integrations as dashed lines. The red dotted line is the sound speed. The bottom panels illustrate the fulfilment of Rankine-Hugoniot jump conditions.}
\label{fig:steady}
\end{figure}

We use these initial values of $v_x$ and $v_y$ to integrate the differential equation both in the backward and forward directions starting from the sonic point $\xs$ (see Fig.~\ref{fig:steady}). Next, we determine if there is a point where the shock jump conditions are satisfied. There are two jump conditions: (i) the component of the velocity parallel to the shock is continuous at the shock, $v_{y,+}=v_{y,-}$; (ii) there must be jump in the perpendicular component such that $v_{x,+}v_{x,-}= \cs^2$. Here, $v_{\pm}$ indicate the velocities just before/after the shock. We first check whether there is a point that quantities, assuming that quantities are periodic with period $L_x$. If there is a point where this condition is satisfied, then we check the second condition. If both conditions are satisfied within a given tolerance ($\sim 10^{-6}$) we stop the procedure and we have found a solution. Otherwise, we change the guess for the sonic point and repeat the procedure until the jump conditions are met.

\section{Initial noise} \label{sec:noise}

Here we discuss in more details the random noise that we introduce to accelerate the onset of the instability and save computational time. We perturb the initial density according to
\begin{equation}
\rho(x,y,t=0) = \rho_0(x)\big(1+h(x,y) \big),\label{4.1}
\end{equation}
where $\rho_0(x)$ is the density of the steady states described in Section~\ref{sec:steady} and $h(x,y)$ is a random noise calculated as follows. We write
\begin{equation}
h(x,y)=\sum_{(k,l)}H_{kl}\exp \bigg[i2\pi \bigg(\frac{xk}{N_x}+\frac{yl}{N_y} \bigg)\bigg], \label{4.2}
\end{equation}
where $N_x$ and $N_y$ are the number of grid points in $x$ and $y$ directions respectively and $k\in \{0,1,2,\cdots N_x-1,N_x \}$, $l\in \{0,1,2,\cdots N_y-1,N_y\}$. We write $H_{kl} = |H_{kl}|e^{i \theta_{kl}}$ and draw the amplitudes $|H_{kl}|$ from a normal distribution with mean $\mu= 0$ and standard deviation $\sigma = 0.01$, and the phases $\theta_{kl}$ from a uniform distribution between $0$ and $2\pi$. In this way we obtain white noise with an amplitue of roughly $|h(x,y)|\simeq 4\%$. The initial noise is visible for example in the initial conditions shown in the top-left corner of Figure~\ref{fig:examples}.

\end{document}